\documentclass[iop]{emulateapj}
\usepackage{natbib}
\usepackage{graphicx}

\newcommand{\cotwo}{\mbox{\rm CO(2-1)}}
\newcommand{\coone}{\mbox{\rm CO(1-0)}}
\newcommand{\hi}{\mbox{\rm H$\,$\scshape{i}}}
\newcommand{\kmpers}{\mbox{km~s$^{-1}$}}
\newcommand{\Kkmpers}{\mbox{K~km~s$^{-1}$}}
\newcommand{\Kkmperspc}{\mbox{K~km~s$^{-1}$~pc$^2$}}

\newcommand{\xcounits}{\mbox{cm$^{-2}$ (K km s$^{-1}$)$^{-1}$}}
\newcommand{\acounits}{\mbox{\rm M$_{\odot}$ pc$^{-2}$} \mbox{(K km s$^{-1}$)$^{-1}$}}
\newcommand{\MJypersr}{\mbox{MJy~sr$^{-1}$}}
\newcommand{\Msunperpc}{\mbox{\rm M$_{\odot}$ pc$^{-2}$}}
\newcommand{\Msun}{\mbox{$M_{\odot}$}}
\newcommand{\Msunperyrperkpc}{\mbox{\rm M$_{\odot}$ yr$^{-1}$ kpc$^{-2}$}}
\newcommand{\Msunperyr}{\mbox{\rm M$_{\odot}$ yr$^{-1}$}}

\shorttitle{CO in Dwarfs}
\shortauthors{Schruba et al.}
\slugcomment{Accepted for publication in \emph{The Astronomical Journal}}

\begin{document}
\title{Low CO Luminosities in Dwarf Galaxies}

\author{
Andreas Schruba\altaffilmark{1},
Adam K.~Leroy\altaffilmark{2},
Fabian Walter\altaffilmark{1},
Frank Bigiel\altaffilmark{3},
Elias Brinks\altaffilmark{4},
W.\,J.\,G. de Blok\altaffilmark{5}, \newline 
Carsten Kramer\altaffilmark{6},
Erik Rosolowsky\altaffilmark{7},
Karin Sandstrom\altaffilmark{1},
Karl Schuster\altaffilmark{8},
Antonio Usero\altaffilmark{9},
Axel Weiss\altaffilmark{10},
Helmut Wiesemeyer\altaffilmark{10}}
\altaffiltext{1}{Max-Planck-Institut f\"ur Astronomie, K\"onigstuhl 17, 69117 Heidelberg, Germany; schruba@mpia.de}
\altaffiltext{2}{National Radio Astronomy Observatory, 520 Edgemont Road, Charlottesville, VA 22903, USA}
\altaffiltext{3}{Zentrum f\"ur Astronomie der Universit\"at Heidelberg, Institut f\"ur Theoretische Astrophysik, Albert-Ueberle-Str.~2, 69120 Heidelberg, Germany}
\altaffiltext{4}{Centre for Astrophysics Research, University of Hertfordshire, Hatfield AL10 9AB, U.K.}
\altaffiltext{5}{Astrophysics, Cosmology and Gravity Centre, Department of Astronomy, University of Cape Town, Private Bag X3, Rondebosch 7701, South Africa}
\altaffiltext{6}{IRAM, Avenida Divina Pastora 7, 18012 Granada, Spain}
\altaffiltext{7}{Department of Physics and Astronomy, University of British Columbia Okanagan, 3333 University Way, Kelowna, BC V1V 1V7, Canada}
\altaffiltext{8}{IRAM, 300 rue de la Piscine, 38406 St.~ŒMartin d\textquoteright H\`{e}res, France}
\altaffiltext{9}{Observatorio Astron\'{o}mico Nacional, Alfonso XII, 3, 28014, Madrid, Spain}
\altaffiltext{10}{MPIfR, Auf dem H\"ugel 69, 53121 Bonn, Germany}

\begin{abstract}
We present maps of $^{12}\text{CO}~J=2-1$ emission covering the entire star-forming disks of 16 nearby dwarf galaxies observed by the IRAM HERACLES survey. The data have 13\arcsec\ angular resolution, $\sim 250$~pc at our average distance of $D=4$~Mpc, and sample the galaxies by $10-1000$ resolution elements. We apply stacking techniques to perform the first sensitive search for CO emission in dwarf galaxies outside the Local Group ranging from individual lines-of-sight, stacking over IR-bright regions of embedded star formation, and stacking over the entire galaxy. We detect 5 galaxies in CO with total CO luminosities of $L_{\rm CO\,2-1} = 3 - 28 \times 10^6$ \Kkmperspc. The other $11$ galaxies remain undetected in CO even in the stacked images and have $L_{\rm CO\,2-1} \lesssim 0.4 - 8 \times 10^6$ \Kkmperspc.   We combine our sample of dwarf galaxies with a large sample of spiral galaxies from the literature to study scaling relations of $L_{\rm CO}$ with $M_{\rm B}$ and metallicity. We find that dwarf galaxies with metallicities of $Z \approx 1/2 - 1/10~Z_\odot$ have $L_{\rm CO}$ of $2-4$ orders of magnitude smaller than massive spiral galaxies and that their $L_{\rm CO}$ per unit $L_{\rm B}$ is $1-2$ orders of magnitude smaller. A comparison with tracers of star formation (FUV and 24\micron) shows that $L_{\rm CO}$ per unit SFR is $1-2$ orders of magnitude smaller in dwarf galaxies. One possible interpretation is that dwarf galaxies form stars much more efficiently, we argue that the low $L_{\rm CO} / \text{SFR}$ ratio is due to the fact that the CO-to-H$_2$ conversion factor, $\alpha_{\rm CO}$, changes significantly in low metallicity environments. Assuming that a constant H$_2$ depletion time of $\tau_{\rm dep} = 1.8$~Gyr holds in dwarf galaxies (as found for a large sample of nearby spirals) implies $\alpha_{\rm CO}$ values for dwarf galaxies with $Z \approx 1/2 - 1/10~Z_\odot$ that are more than one order of magnitude higher than those found in solar metallicity spiral galaxies. Such a significant increase of $\alpha_{\rm CO}$ at low metallicity is consistent with previous studies, in particular those of Local Group dwarf galaxies which model dust emission to constrain H$_2$ masses. Even though it is difficult to parameterize the dependence of $\alpha_{\rm CO}$ on metallicity given the currently available data the results suggest that CO is increasingly difficult to detect at lower metallicities. This has direct consequences for the detectability of star-forming galaxies at high redshift which presumably have on average sub-solar metallicity.
\end{abstract}

\keywords{galaxies: ISM --- ISM: molecules --- radio lines: galaxies}

\section{Introduction}
\label{introduction}

Robust knowledge of the molecular (H$_2$) gas distribution is indispensable to understand star formation in galaxies. Observations in the Milky Way and nearby galaxies suggest that stars form in clouds consisting predominantly of H$_2$ \citep{Lada2003, Fukui2010}. Because H$_2$ is almost impossible to observe directly under typical conditions of the cold interstellar medium (ISM), its abundance and distribution has to be inferred using indirect methods. Observations of low rotational lines of carbon monoxide (CO) have been the standard method to do so as CO is the second most abundant molecule and easily excited in the cold ISM. Over the last decades, of the order of a hundred galaxies in the local Universe have been successfully detected in CO. Over the last years CO has been detected throughout the Universe out to cosmological distances \citep{Solomon2005}. These CO observations have greatly enhanced our knowledge of H$_2$ in galaxies, the phase balance of the ISM, and its interplay with star formation.

Despite great advances in studying H$_2$ in massive star-forming galaxies, our knowledge of H$_2$ in star-forming dwarf galaxies remains poor. The CO emission in these systems has proven to be extremely faint and most studies targeting metal-poor dwarf galaxies have resulted in non-detections. For sensitivity reasons, surveys of dwarf galaxies have tended to target only a few systems and used mostly single pointings \citep{Israel1995, Young1995, Taylor1998, Barone2000, Boker2003, Sauty2003, Albrecht2004, Leroy2005}. These data are very heterogeneous as they target different CO transitions, cover different regions, and have different beam sizes, sensitivities, and beam filling factors. Thus, conclusive results for basic quantities such as the total CO luminosity of dwarf galaxies have not been reached and comparison to other observables have been complicated by these systematic effects.

CO observations are currently --- and will remain --- our most accessible tracer of cold H$_2$ in the local and distant Universe. It is thus important to obtain profound understanding of the connection between CO and H$_2$ in different environments. Inside individual molecular clouds, this dependence has proven to be highly complicated and influenced by many factors \citep[e.g.,][]{Shetty2011a, Shetty2011b}. Many of these dependencies average out on scales larger than individual clouds, however metallicity will not. Metallicity may thus be the single most important factor determining the CO/H$_2$ ratio on large scales. This makes a robust calibration of the CO/H$_2$ ratio as function of metallicity a viable proposition for those observational studies that use CO as a tracer of H$_2$. The need becomes more pressing as observations start probing the CO content of galaxies in the distant Universe where most stars presumably formed in environments with sub-solar metallicity.

To understand the environmental dependencies of the CO/H$_2$ ratio requires good knowledge of the CO content of all types of galaxies, even in those where we worry that CO may not trace H$_2$ in the same way as it does in massive spiral galaxies. This makes sensitive, wide-field CO maps of dwarf galaxies an important undertaking. The HERACLES\footnote{\url{http://www.cv.nrao.edu/~aleroy/HERACLES/Overview.html}\label{foot1}} survey \citep[partly published in][]{Leroy2009a} has obtained such CO observations of a large set of nearby star-forming galaxies ranging from massive spirals down to low-mass, low-metallicity dwarfs. In conjunction with an extensive set of multi-wavelength data, this survey has already led to a vast improvement of our knowledge of the relation between \hi, CO, H$_2$, and star formation. 

In this paper, we present sensitive measurements of CO emission of 16 nearby low-mass, low-metallicity star-forming dwarf galaxies from the HERACLES survey using stacking techniques. We use these data to study the relation between CO emission and other galaxy parameters, especially star formation rate (SFR) and H$_2$ mass. Then we analyze the CO/H$_2$ ratio as function of metallicity. In Section~\ref{data} we introduce our multi-wavelength data and summarize their basic properties. In Section~\ref{coemission} we conduct a sensitive search for CO emission for individual lines-of-sight, IR-bright regions, and entire galaxies. In Section~\ref{corelations} we compare these CO measurements to other galaxy parameters and compare the relationships found for dwarf galaxies to those of massive spiral galaxies. In Section~\ref{xcofactor} we study the metallicity dependence of the CO/H$_2$ ratio. We use observed SFRs to infer H$_2$ masses and thus constrain CO/H$_2$, then we compare our results to results derived from other methods.  In Section~\ref{summary} we summarize our findings.

\section{Data}
\label{data}

\begin{figure*}
\epsscale{0.85} \plotone{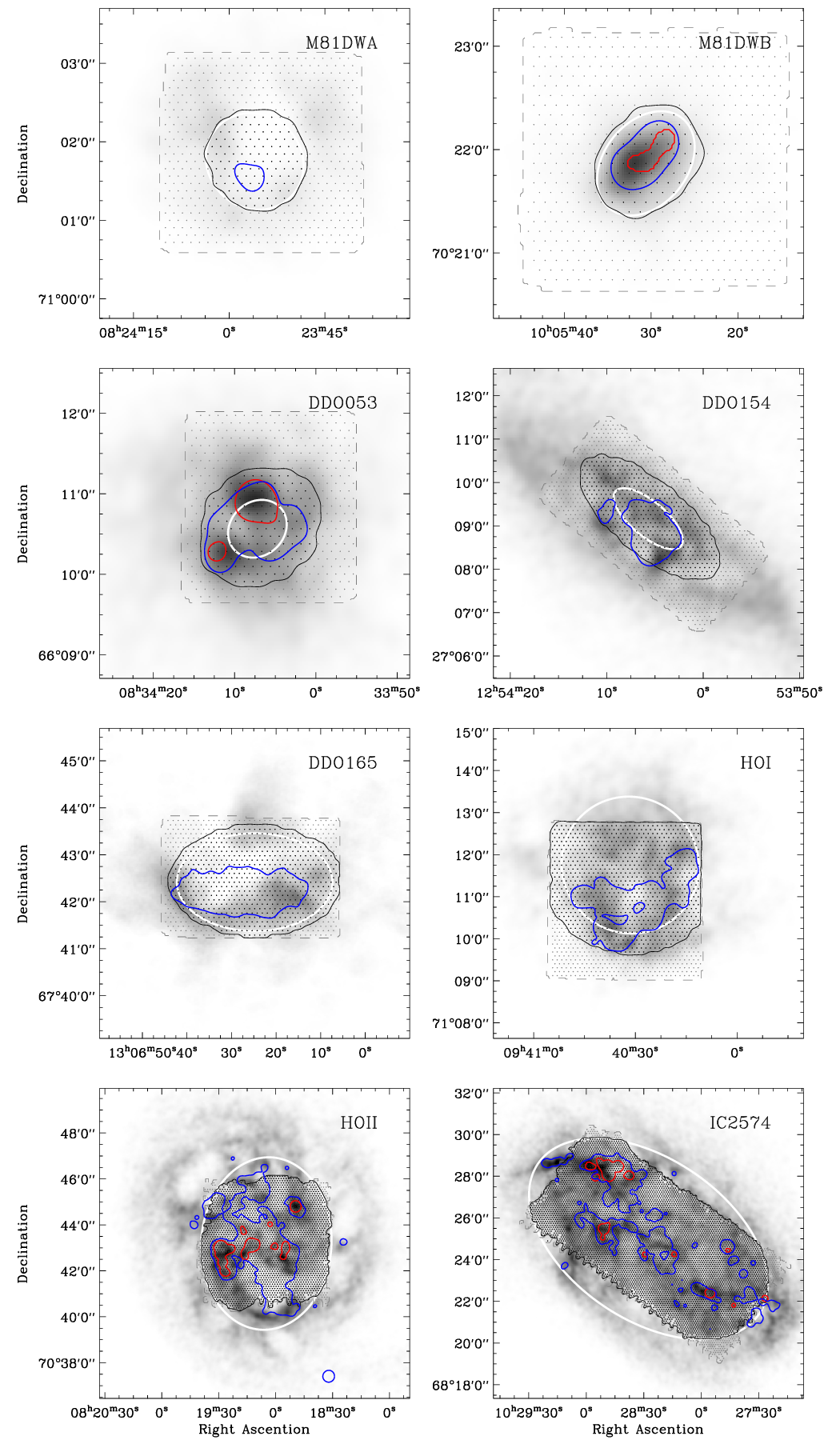}
\caption{{\em Coverage of CO Observations.} For each HERACLES dwarf galaxy we show the coverage of our CO data (gray dashed line), the \hi\ surface density (grayscale) at linear scale between $0-40$ \Msunperpc, the 24\micron\ intensity (red contour) at $0.2$ \MJypersr, the FUV intensity (blue contour) at $0.01$ \MJypersr, and a galactocentric radius $R = R_{25}$ (white contour). We determine the CO intensity at each sampling point (dot), and stack the data for the ``entire'' galaxy (black contour) and for IR-bright regions (red contour), see text for definition of stacking regions.\label{f1}}
\end{figure*}

\addtocounter{figure}{-1}
\begin{figure*}
\epsscale{0.85} \plotone{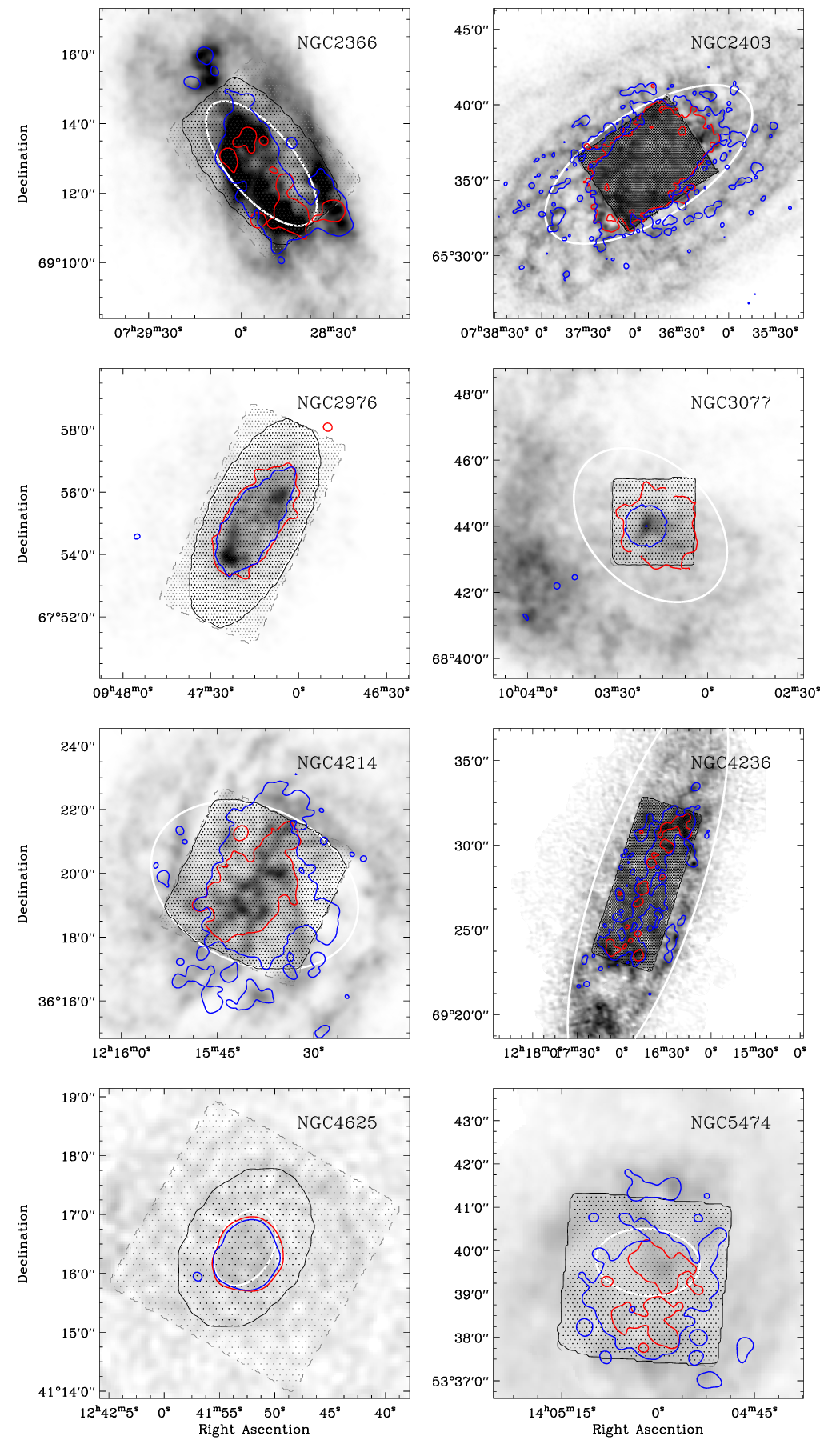}
\caption{(Continued)\label{f1}}
\end{figure*}

\begin{deluxetable*}{l*{9}{c}}
\tablecolumns{10}
\tablecaption{Properties of Galaxy Sample\label{t1}}
\tablehead{\colhead{Name} & \colhead{Alt.~Name} & \colhead{$D$} & \colhead{Incl.} & \colhead{P.A.} & \colhead{$R_{\rm 25}$} & \colhead{Metal.\tablenotemark{a}} & \colhead{$M_{\rm B}$\tablenotemark{b}} & \colhead{$M_{\rm HI}$\tablenotemark{c}} & \colhead{SFR\tablenotemark{d}} \\
\colhead{} & \colhead{} & \colhead{(Mpc)} & \colhead{($^{\circ}$)} & \colhead{($^{\circ}$)} & \colhead{($\arcmin$)} & \colhead{12+logO/H} & \colhead{(mag)} & \colhead{($10^8~M_\odot$)} & \colhead{($M_\odot$~yr$^{-1}$)}}
\startdata
M\,81 Dw\,A &   & 3.6 & 23 & 49 & 0.64 & 7.50 & -11.4 & 0.12 & 0.0005 \\
M\,81 Dw\,B & UGC 5423 & 5.3 & 44 & 321 & 0.56 & 8.02 & -13.8 & 0.25 & 0.0023 \\
DDO 053 & UGC 4459 & 3.6 & 31 & 132 & 0.39 & 7.80 & -13.9 & 0.60 & 0.0035 \\
DDO 154 & UGC 8024 & 4.3 & 66 & 230 & 0.98 & 7.78 & -15.4 & 3.58 & 0.0056 \\
DDO 165 & UGC 8201 & 4.6 & 51 & 90 & 1.66 & 7.84 & -14.1 & 6.33 & 0.0100 \\
HO I & UGC 5139 & 3.8 & 12 & 50 & 1.65 & 7.83 & -16.8 & 1.39 & 0.0100 \\
HO II & UGC 4305 & 3.4 & 41 & 177 & 3.76 & 7.93 & -12.5 & 5.95 & 0.0455 \\
IC 2574 & UGC 5666 & 4.0 & 53 & 56 & 6.41 & 8.05 & -17.2 & 14.80 & 0.0718 \\
NGC 2366 & UGC 3851 & 3.4 & 64 & 40 & 2.20 & 7.96 & -16.2 & 6.49 & 0.0605 \\
NGC 2403 & UGC 3918 & 3.2 & 63 & 124 & 7.87 & 8.57 & -18.6 & 25.80 & 0.4140 \\
NGC 2976 & UGC 5221 & 3.6 & 65 & 335 & 3.60 & 8.67 & -16.5 & 1.36 & 0.0895 \\
NGC 3077 & UGC 5398 & 3.8 & 46 & 45 & 2.70 & 8.64 & -17.3 & 8.81 & 0.0838 \\
NGC 4214 & UGC 7278 & 2.9 & 44 & 65 & 3.40 & 8.25 & -17.1 & 4.08 & 0.1208 \\
NGC 4236 & UGC 7306 & 4.4 & 75 & 162 & 11.99 & 8.46 & -18.1 & 34.60 & 0.1409 \\
NGC 4625 & UGC 7861 & 9.5 & 47 & 330 & 0.69 & 8.70 & -17.0 & 11.80 & 0.0716 \\
NGC 5474 & UGC 9013 & 6.8 & 50 & 85 & 1.20 & 8.57 & -17.3 & 15.50 & 0.1069
\enddata
\tablenotetext{a}{Oxygen abundance from \citet{Moustakas2010}.}
\tablenotetext{b}{$B$-band magnitude from HERACLES.}
\tablenotetext{c}{$M_{\rm HI}$ from \citet{Walter2008}.}
\tablenotetext{d}{SFR(FUV+24) from this work.}
\end{deluxetable*}

We study 16 nearby low-mass star-forming galaxies (for simplicity just called ``dwarfs'' throughout the paper) from the HERACLES\footnotemark[\value{footnote}] survey \citep{Leroy2009a}; see Figure~\ref{f1} for an outline of the area surveyed for each target. These data have been largely neglected in previous work on the HERACLES sample as their CO emission has rarely been robustly detected in the pixel-based studies of \citet{Bigiel2008,Bigiel2011} and \citet{Leroy2008} or the radial stacking analysis of \citet{Schruba2011}. The only galaxies for which data were provided are Ho~I, Ho~II, DDO 154, IC 2574, NGC 2976, and NGC 4214 \citep{Leroy2009a}. Table~\ref{t1} lists our sample of dwarf galaxies along with adopted distances, inclination, position angle, optical radius, metallicity, $B$-band optical magnitude, \hi\ mass, and total star formation rate (SFR). These values are taken from \citet{Walter2008} where possible and from LEDA \citep{Prugniel1998} and NED elsewhere.

\subsection{CO Data}
\label{codata}

We take CO data from the HERACLES\footnotemark[\value{footnote}] survey which mapped the $^{12}$CO \mbox{$J=2\rightarrow1$} emission line in 48 nearby galaxies using the IRAM 30m telescope \citep{Leroy2009a}. The observations are designed to cover large parts of the galaxies and extend to $1-1.5$ times the optical radius, $R_{25}$, for large spirals and up to $2-3~R_{25}$ for small galaxies. The final data cubes have an angular resolution (FWHM) of 13\arcsec\ and a spectral resolution (channel separation) of $2.6$~\kmpers. The noise level is $20 - 30$~mK per resolution element and per channel.

Whenever possible we compare our observed \cotwo\ data to literature measurements. To do that, we convert them to \coone\ intensities assuming a constant line ratio, $R_{21} = I_{\rm CO~2-1}~/$ $I_{\rm CO~1-0} = 0.7$. We choose this constant line ratio to achieve consistency with \citet{Bigiel2011} and \citet{Schruba2011}. This is the average ratio found for all HERACLES galaxies (E.~Rosolowsky et al., in preparation). From this data no significant variations of $R_{21}$ with metallicity are evident over the range $12+\textrm{log\,O/H} \approx 8.6 - 8.9$ on scales of 1~kpc. In low metallicity environments such as the Magellanic clouds values of $R_{21} \sim 1.0 - 1.5$ are frequently found \citep{Bolatto2000, Bolatto2003, Israel2003, Israel2005}. This can lead to an over-prediction of the true \coone\ values by a factor of $\lesssim 2$. However, later we will see that this potential bias is too small to change our conclusions.

We will discuss the CO-to-H$_2$ conversion factor extensively in Section~\ref{xcofactor} but note that a typical Galactic \mbox{CO(1-0)}-to-H$_2$ conversion factor is $X_{\rm CO} = 2.0 \times 10^{20}$ \xcounits\ \citep[][]{Strong1996, Dame2001, Abdo2010} which translates to $\alpha_{\rm CO} = 4.38$ \acounits\ when including a factor of $1.36$ to account for heavy elements.

\subsection{HI Data}
\label{hidata}

We draw \hi\ data mostly from the VLA THINGS survey \citep{Walter2008}. The \hi\ data for NGC 4236, NGC 4626, and NGC 5474 are from the VLA programs AL~731, AL~735, both led by one of us (P.I.~Leroy), and from the archive. The \hi\ data for DDO 165 is from the LITTLE THINGS survey (D.~Hunter et al., in preparation). The angular resolution of the data referred to above is $\sim 10 - 20$\arcsec, the velocity resolution is $2.6 - 5.2$ \kmpers\ for the THINGS and LITTLE THINGS data, and $5.2 - 10$ \kmpers\ for the other data. The sensitivity of these data is sufficiently high to never limit our analysis.

\subsection{Star Formation Tracers}
\label{sftracers}

We estimate the star formation rate (SFR) using a combination of FUV and 24\micron\ emission following the approach introduced in \citet{Bigiel2008} and \citet{Leroy2008}. The SFR surface density is given by $\Sigma_{\rm SFR}$ [\Msunperyrperkpc] = $0.081~(I_{\rm FUV} + 0.04~I_{\rm 24\mu m})$ [\MJypersr] $\times~\text{cos}~i$. The FUV data are taken from the {\em GALEX} Nearby Galaxy Survey \citep{GildePaz2007} or alternatively from the NASA Multimission Archive at STScI. They cover a wavelength range of $1350 - 1750$~\AA, have angular resolution $\sim 4.5$\arcsec, and sufficient sensitivity to determine FUV intensities with high signal-to-noise throughout the star-forming disk. The IR data are taken from the {\em Spitzer} SINGS \citep{Kennicutt2003b} and Local Volume Legacy (LVL) surveys \citep{Dale2009}. These data have $\sim 6$\arcsec\ resolution; their sensitivity is sufficient to detect 24\micron\ emission in most of our galaxies except the lowest mass and lowest metallicity dwarf galaxies. We apply some processing to the FUV and 24\micron\ maps (i.e., mask foreground stars and flatten background) as described in \citet{Leroy2012}.

\subsection{Metallicities}
\label{metallicities}

Gas phase oxygen abundances (metallicities) are taken from \citet{Moustakas2010}. For galaxy-integrated data we use the average of their ``characteristic'' metallicities (their Table~9) derived from a theoretical calibration (KK04 values) and an empirical calibration (PT05 values). In some plots we also show the radial stacking results from \citet{Schruba2011} for which we use the metallicity gradients from \citet[][Table~8]{Moustakas2010} again averaging the two calibrations. The ``characteristic'' metallicity of a galaxy corresponds to the value of the metallicity gradient at radius $R = 0.4~R_{25}$.

\subsection{Sampling}
\label{sampling}

We convolve all our data to a common resolution of 13\arcsec\ (limited by the CO data; some \hi\ data have coarser beam sizes and we include them on their native resolution assuming to first order homogeneous \hi\ distribution) and sample them on a hexagonally packed grid spaced by half a beam size ($6.5$\arcsec). For each line-of-sight we collect observed intensities of CO, \hi, FUV, and 24\micron, and determined local gas masses and SFRs. We also store the \hi\ mean velocity, the original CO spectrum, and the galactocentric radius. Figure~\ref{f1} shows for each galaxy in our sample the \hi\ distribution as grayscale, the extent of FUV and 24\micron\ emission indicated by a single contour, the CO map coverage, and our sampling grid as dots. See Figure~\ref{f2} for integrated CO intensity maps for a subset of our galaxies.

\subsection{Literature Sample}
\label{literaturesample}

\begin{deluxetable*}{l*{10}{c}}
\tablecolumns{11}
\tablecaption{Properties of Literature Galaxy Sample\label{t2}}
\tablehead{\colhead{Name} & \colhead{$D$} & \colhead{Ref.\tablenotemark{a}} & \colhead{Metal.} & \colhead{Ref.\tablenotemark{a}} & \colhead{$M_{\rm B}$} & \colhead{Ref.\tablenotemark{a}} & \colhead{log $L_{\rm CO~1-0}$} & \colhead{Ref.\tablenotemark{a}} & \colhead{log SFR} & \colhead{Ref.\tablenotemark{a}} \\
\colhead{} & \colhead{(Mpc)} & \colhead{} & \colhead{12+logO/H} & \colhead{} & \colhead{(mag)} & \colhead{} & \colhead{(K~km~s$^{-1}$~pc$^2$)} & \colhead{} & \colhead{($M_\odot$~yr$^{-1}$)} & \colhead{}}
\startdata
SMC & 0.06 & L11 & 8.00 & D84; MA10 & -16.2 & L11 & 5.20 & M06 & -1.30 & W04 \\
LMC & 0.05 & L11 & 8.30 & D84; MA10 & -17.6 & L11 & 6.50 & F08 & -0.70 & H09 \\
IC 10 & 0.95 & H01 & 8.20 & L79; L03 & -16.5 & H01 & 6.30 & L06 & -1.03 & L06 \\
M 33 & 0.84 & G04 & 8.30 & R08 & -18.9 & NED & 7.60 & H04 & 0.00 & H04 \\
I ZW 18 & 14.00 & I04 & 7.22 & T05 & -14.7 & G03 & $<$ 0.10 & L07 & -1.00 & L07 \\
II ZW 40 & 9.20 & C10 & 8.10 & E08; C10 & -17.9 & NED & 6.20 & T98 & -0.19 & C10 \\
NGC 0628 & 7.30 & W08 & 8.69 & MO10 & -20.0 & W08 & 8.45 & HERA & -0.08 & HERA \\
NGC 0925 & 9.20 & W08 & 8.52 & MO10 & -20.0 & W08 & 7.47 & HERA & -0.24 & HERA \\
NGC 1482 & 22.00 & C10 & 8.53 & MO10 & -18.8 & NED & 8.80 & Y95 & 0.53 & C10 \\
NGC 1569 & 3.36 & G08 & 8.10 & M97 & -18.1 & NED & 5.55 & T98 & -0.40 & P11 \\
NGC 2146 & 12.80 & W08 & 8.70 & E08; C10 & -20.6 & W08 & 9.06 & HERA & 0.93 & C10 \\
NGC 2537 & 6.90 & L11 & 8.40 & MA10 & -16.4 & L11 & 5.50 & T98 & -1.05 & C10 \\
NGC 2782 & 40.00 & C10 & 8.60 & E08; C10 & -20.9 & NED & 9.00 & Y95 & 0.72 & C10 \\
NGC 2798 & 24.70 & W08 & 8.69 & MO10 & -19.4 & W08 & 8.73 & HERA & 0.49 & C10 \\
NGC 2841 & 14.10 & W08 & 8.88 & MO10 & -21.2 & W08 & 8.34 & HERA & -0.10 & HERA \\
NGC 2903 & 8.90 & W08 & 8.90 & MA10 & -20.1 & L11 & 8.82 & HERA & 0.32 & HERA \\
NGC 3034 & 3.60 & W08 & 8.82 & MO10 & -18.5 & L11 & 8.94 & HERA & 0.90 & C10 \\
NGC 3079 & 21.80 & C10 & 8.60 & E08; C10 & -21.7 & NED & 9.40 & Y95 & 0.50 & C10 \\
NGC 3184 & 11.10 & W08 & 8.83 & MO10 & -19.9 & W08 & 8.56 & HERA & -0.01 & HERA \\
NGC 3198 & 13.80 & W08 & 8.62 & MO10 & -20.7 & W08 & 8.15 & HERA & -0.01 & HERA \\
NGC 3310 & 21.30 & C10 & 8.20 & E08; C10 & -20.5 & NED & 8.20 & Y95 & 0.92 & C10 \\
NGC 3351 & 10.10 & W08 & 8.90 & MO10 & -19.5 & L11 & 8.41 & HERA & -0.01 & HERA \\
NGC 3368 & 10.52 & L11 & 9.00 & MA10 & -20.0 & L11 & 8.30 & Y95 & -0.45 & C10 \\
NGC 3521 & 10.70 & W08 & 8.70 & MO10 & -20.3 & L11 & 8.96 & HERA & 0.34 & HERA \\
NGC 3627 & 9.30 & W08 & 8.66 & MO10 & -20.1 & L11 & 8.84 & HERA & 0.36 & HERA \\
NGC 3628 & 9.40 & L11 & 9.00 & MA10 & -19.6 & L11 & 9.20 & Y95 & 0.33 & C10 \\
NGC 3938 & 12.20 & W08 & 8.70 & E08; C10 & -19.6 & W08 & 8.41 & HERA & -0.07 & HERA \\
NGC 4194 & 42.00 & C10 & 8.70 & E08; C10 & -20.5 & NED & 8.90 & Y95 & 1.13 & C10 \\
NGC 4254 & 20.00 & W08 & 8.79 & MO10 & -21.3 & W08 & 9.50 & HERA & 0.83 & HERA \\
NGC 4321 & 14.30 & W08 & 8.84 & MO10 & -20.9 & W08 & 9.20 & HERA & 0.45 & HERA \\
NGC 4449 & 4.20 & W08 & 8.30 & M97 & -18.1 & L11 & 7.01 & B03 & -0.45 & C10 \\
NGC 4450 & 27.10 & C10 & 8.90 & C10; MA10 & -21.7 & NED & 8.90 & Y95 & -0.18 & C10 \\
NGC 4536 & 14.50 & W08 & 8.60 & MO10 & -19.7 & W08 & 8.60 & HERA & 0.42 & HERA \\
NGC 4569 & 20.00 & W08 & 8.90 & E08; C10 & -22.1 & W08 & 9.14 & HERA & 0.29 & HERA \\
NGC 4579 & 20.60 & W08 & 9.00 & C10; MA10 & -21.4 & W08 & 8.94 & HERA & 0.11 & HERA \\
NGC 4631 & 8.90 & W08 & 8.43 & MO10 & -19.9 & L11 & 8.72 & HERA & 0.40 & C10 \\
NGC 4725 & 9.30 & W08 & 8.73 & MO10 & -20.2 & W08 & 7.85 & HERA & -0.43 & HERA \\
NGC 4736 & 4.70 & W08 & 8.66 & MO10 & -19.4 & L11 & 8.14 & HERA & -0.29 & HERA \\
NGC 4826 & 7.50 & W08 & 8.87 & MO10 & -20.0 & L11 & 8.10 & H03 & -0.50 & C10 \\
NGC 5033 & 14.80 & MO10 & 8.66 & MO10 & -20.8 & NED & 9.30 & H03 & 0.10 & K03 \\
NGC 5055 & 10.10 & W08 & 8.77 & MO10 & -20.7 & L11 & 9.10 & HERA & 0.34 & HERA \\
NGC 5194 & 8.00 & W08 & 8.86 & MO10 & -20.6 & L11 & 9.20 & HERA & 0.49 & HERA \\
NGC 5236 & 4.50 & W08 & 9.00 & MO10 & -20.1 & L11 & 8.90 & Y95 & 0.37 & C10 \\
NGC 5253 & 3.15 & L11 & 8.20 & MA10 & -16.6 & L11 & 5.80 & T98 & -0.22 & C10 \\
NGC 5713 & 26.50 & W08 & 8.64 & MO10 & -20.9 & W08 & 9.17 & HERA & 0.76 & HERA \\
NGC 5866 & 15.10 & C10 & 8.70 & C10; MA10 & -20.2 & NED & 8.10 & Y95 & -0.60 & C10 \\
NGC 5953 & 35.00 & C10 & 8.70 & E08; C10 & -20.0 & NED & 9.00 & Y95 & 0.38 & C10 \\
NGC 6822 & 0.49 & G10 & 8.11 & LE06 & -15.2 & NED & 5.15 & G10 & -1.85 & E11 \\
NGC 6946 & 5.90 & W08 & 8.73 & MO10 & -19.2 & L11 & 9.04 & HERA & 0.57 & HERA \\
NGC 7331 & 14.70 & W08 & 8.68 & MO10 & -21.7 & NED & 9.10 & HERA & 0.49 & HERA
\enddata
\tablenotetext{a}{References: B03 = \citet{Bottner2003}; C10 = \citet{Calzetti2010}; D84 = \citet{Dufour1984}; E08 = \citet{Engelbracht2008}; E11 = \citet{Efremova2011}; F08 = \citet{Fukui2008}; G03: \citet{GildePaz2003}; G04: \citet{Galleti2004}; G08 = \citet{Grocholski2008}; G10 = \citet{Gratier2010a}; H01 = \citet{Hunter2001}; H03 = \citet{Helfer2003}; H04 = \citet{Heyer2004}; H09 = \citet{Harris2009}; I97 = \citet{Israel1997}; I04 = \citet{Izotov2004}; K03 = \citet{Kennicutt2003b}; L79 = \citet{Lequeux1979}; L03 = \citet{Lee2003}; LE06 = \citet{Lee2006}; L06 = \citet{Leroy2006}; L07 = \citet{Leroy2007}; L11 = \citet{Lee2011}; M97 = \citet{Martin1997}; M06 = \citet{Mizuno2006}; MA10 = \citet{Marble2010}; MO10 = \citet{Moustakas2010}; P11 = \citet{Pasquali2011}; R08 = \citet{Rosolowsky2008}; T98 = \citet{Taylor1998}; T05 = \citet{Thuan2005}; W04 = \citet{Wilke2004}; W08 = \citet{Walter2008}; Y95 = \citet{Young1995}; HERA = HERACLES collaboration.}
\end{deluxetable*}

Throughout the paper we will compare our measurements for dwarf galaxies to a larger sample of nearby galaxies. This sample is taken from the literature compilation of \citet{Krumholz2011} which includes the more massive HERACLES galaxies and some additional Local Group and nearby galaxies. Table~\ref{t2} lists their names together with adopted distances, metallicities, $B$-band magnitudes, total CO\mbox(1-0) luminosity, total SFR, and references to the original literature. The compilation aims at maximizing homogeneity of used data and methodology. We supplement the \citeauthor{Krumholz2011} compilation by adding absolute $B$-band magnitudes adjusted to our adopted distances. We also update the total CO luminosities using the most recent HERACLES data (converted to CO(1-0) luminosities) and SFRs derived from combining FUV and 24\micron\ maps. We use the metallicities listed in \citeauthor{Krumholz2011}, these have been derived following the above described methodology.

\section{CO Emission in HERACLES Dwarf Galaxies}
\label{coemission}

\begin{figure*}
\epsscale{0.85} \plotone{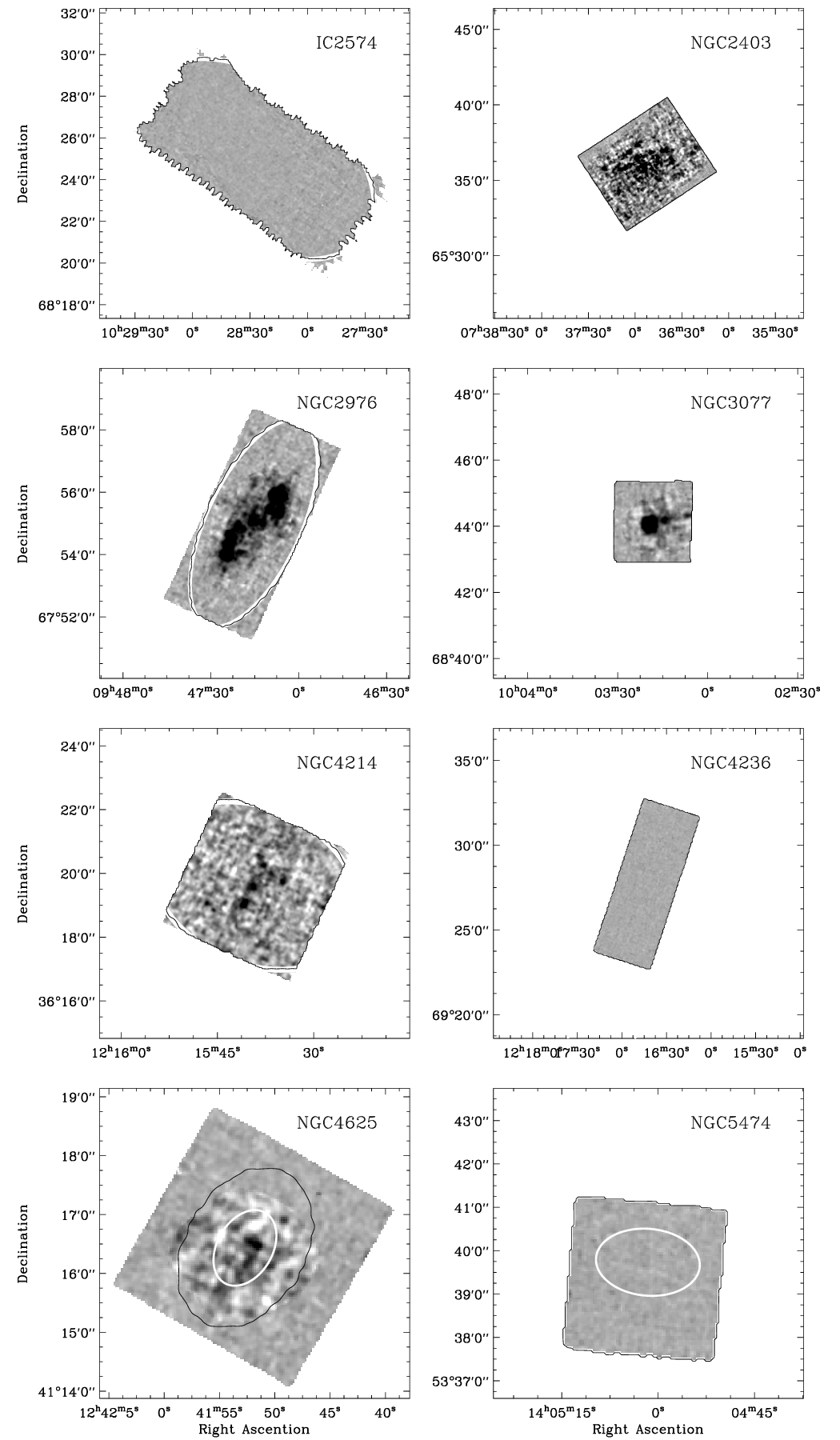}
\caption{{\em CO Intensity Maps.} Integrated \cotwo\ intensity maps (including all channels with velocities in the range of $\pm 50$ \kmpers\ of the local \hi\ mean velocity) for a subsample of the HERACLES dwarf galaxies at a linear grayscale from $-0.3$ to $0.7$ \Kkmpers. The galactocentric radius $R = R_{25}$ is shown as white contour, the region selected to determine the integrated CO intensity for the entire galaxy as black contour. The remaining HERACLES dwarf galaxies not shown here are non-detections at full resolution.\label{f2}}
\end{figure*}

To derive meaningful constraints on CO content, we search for CO emission on three different spatial scales: individual lines-of-sight, stacked over the entire galaxy (i.e., map coverage), and stacked over regions bright in 24\micron.

\subsection{Individual Lines of Sight}
\label{sensitivity}

\begin{deluxetable}{lcl}
\tablecolumns{3}
\tablecaption{CO Luminosities of Molecular Clouds\label{t3}}
\tablehead{\colhead{Name} & \colhead{$L_{\rm CO~1-0}$} & \colhead{Reference} \\
\colhead{} & \colhead{(\Kkmperspc)} & \colhead{}}
\startdata
M\,33 EPRB 1 & $1.8 \times 10^5$ & \citet{Rosolowsky2003} \\
LMC N\,197 & $7.0 \times 10^5$ & \citet{Fukui2008} \\
SMC N\,84 & $1.3 \times 10^4$ & \citet{Mizuno2001} \\
IC\,10 B11a & $7.6 \times 10^4$ & \citet{Leroy2006} \\
Orion-Monoceros & $8.6 \times 10^4$ & \citet{Wilson2005} \\
Orion A & $2.7 \times 10^4$ & \citet{Wilson2005} \\
Taurus & $5.6 \times 10^3$ & \citet{Goldsmith2008}
\enddata
\end{deluxetable}

We start with searching for significant CO emission in individual lines-of-sight. For the dwarf galaxies in HERACLES the noise per channel map in the full resolution ($13\arcsec \times 2.6$ \kmpers) cubes is $\sigma = 21 \pm 3$~mK. For each galaxy we search the entire cube for regions with signal-to-noise ratio (SNR) $>4$ over two consecutive velocity channels. This corresponds to a CO point source with luminosity $L_{\rm CO~2-1} = 2.1 \times 10^4~(\sigma / 20~{\rm mK})~(D / 4~{\rm Mpc})^2$ \Kkmperspc. For comparison, Table~\ref{t3} lists CO(1-0) luminosities of the brightest clouds in M~33, LMC, SMC, IC~10, and values for the Milky Way Orion-Monoceros complex, Orion~A, and Taurus. We are sensitive enough to detect these clouds (except Taurus) at our average source distance of $D = 4$~Mpc.

Figure~\ref{f2} shows maps of integrated CO intensity for the more massive dwarf galaxies of our sample. Each line-of-sight integral includes all channels with velocities within the range of $\pm 50$ \kmpers\ of the local \hi\ mean velocity. Five galaxies, NGC 2403, NGC 2976, NGC 3077, NGC 4214,  and NGC 4625, show emission exceeding our point source sensitivity within $\pm 50$~\kmpers\ of the local mean \hi\ velocity. A point source of $1.5$ times our point source sensitivity will show up completely black at the chosen linear grayscale. For all other galaxies we detect no signal at this angular resolution. The non-detection of bright CO clouds in most of our targets is most likely linked to their low metallicity, $12+\text{log\,O/H} \lesssim 8.0$, while the reference sample in Table~\ref{t3} has higher metallicities, $8.2 \lesssim 12+\text{log\,O/H} \lesssim 8.8$.

\subsection{Improve Sensitivity by Stacking}
\label{stacking}

\begin{figure*}
\epsscale{0.85} \plotone{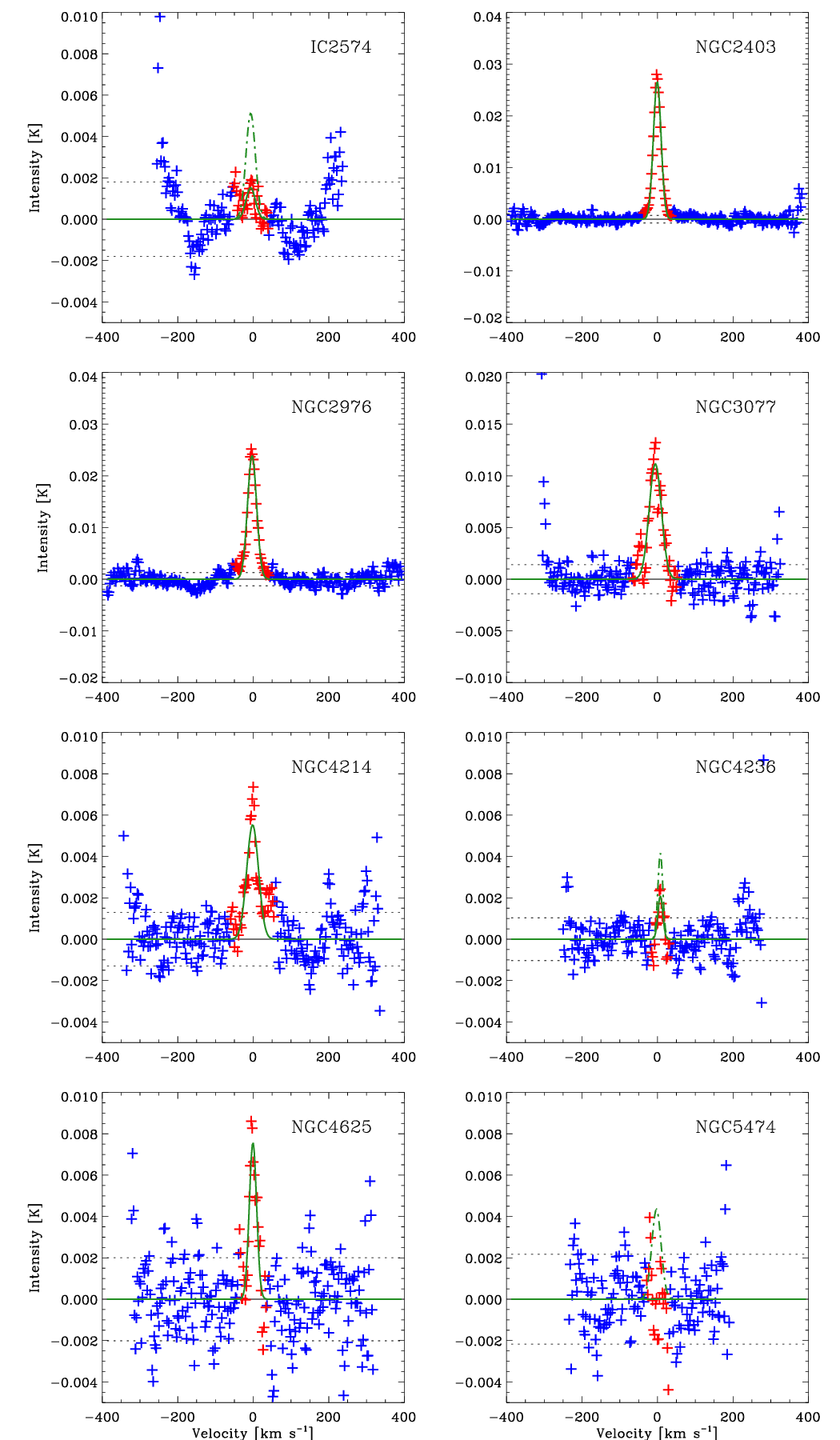}
\caption{{\em Stacked CO Spectra.} The resulting mean \cotwo\ spectrum after stacking all data over the entire galaxy (or map extent; black contours in Figures \mbox{\ref{f1} \& \ref{f2}}). The spectra are shifted to the local mean \hi\ velocity and thus expected to peak at zero velocity (see text). We fit Gaussian profiles (green lines) to the data within $\pm 50$ \kmpers\ to determine the integrated CO intensity; for stacked spectra without robust signal we determine 3$\sigma$ upper limit (green dot-dashed lines). The horizontal dotted lines show the 1$\sigma$ rms noise of the stacked spectra.\label{f3}}
\end{figure*}

The large map size of the HERACLES maps and the fine (13\arcsec) resolution as compared to the angular extent of the galaxies allow us to search for CO emission at many different locations inside the galaxies. We saw above that only a few galaxies have signal strong enough to be detected in individual lines-of-sight. Therefore, we now apply the stacking technique developed and described in detail in \citet{Schruba2011}. This method accounts for the velocity shift in the observed CO spectrum due to galaxy rotation or other bulk motion. This is done by re-adjusting the velocity axis of the CO spectrum of each line-of-sight such that the local \hi\ mean velocity appears at a common (zero) velocity in the shifted spectrum. Under the assumption that the mean velocities of \hi\ and CO closely correspond to each other (confirmed in the bright inner disk of spiral galaxies), the CO line peaks in the shifted spectra by construction at zero velocity across each galaxy (and across the sample). By averaging these shifted spectra we can decrease the noise and coherently add up the spectral line at known (zero) velocity. We may expect that the signal-to-noise (SNR) in the integrated line intensity, $L_{\rm CO}$, does improve proportional to $\sqrt{\Delta V_{\rm Gal}} / \sqrt{\Delta V_{\rm Obs}}$, where $\Delta V_{\rm Gal}$ is the total velocity gradient across the galaxy from galaxy rotation and $\Delta V_{\rm Obs}$ is the width of the CO line at the observing resolution. This ratio may typically be on the order of $\sqrt{250} / \sqrt{20} \sim 3.5$. Significant further improvement in the SNR will be achieved by averaging over many lines-of-sight.

To determine the line intensity we fit the stacked spectrum by a Gaussian profile with center restricted to be within $\pm 50$~\kmpers\ of zero velocity, FWHM to be larger than $15$~\kmpers, and the amplitude to be positive. In cases where the fitted Gaussian has peak intensity below 3$\sigma$ or the integrated intensity is less than 5 times its uncertainty we determine an upper limit instead. The upper limit is defined as the integrated intensity of a Gaussian profile with FWHM set to $18$~\kmpers\ and amplitude fixed to 3$\sigma$. Figure~\ref{f3} shows stacked CO spectra determined over the entire galaxy for the targets shown in Figure~\ref{f2}.

For white noise $\sigma_{\rm rms}$ (in the velocity-integrated intensity of the stacked spectrum) decreases proportional to $N^{-1/2}$ by stacking where $N$ is the number of independent resolution elements. For our data $\sigma_{\rm rms}$ does improve by stacking but at somewhat slower rate and saturates at  $\sim 1$~mK~\kmpers\ after averaging over $\sim 500 - 1000$ resolution elements. Deviations from the white noise behavior are linked to our observing strategy and data reduction \citep[see][]{Leroy2009a}.

It is further instructive to note that the size of the selected stacking region becomes a critical quantity as we have to deal with non-detections. Increasing the size of the stacking region will lead to stronger upper limits on $I_{\rm CO}$; in an ideal world $I_{\rm CO} \propto N^{-1/2}$. The upper limits on $L_{\rm CO}$ will however degrade with increasing area since $L_{\rm CO} = I_{\rm CO} \times Area \propto N^{1/2}$. As we are mainly interested in the absolute quantity $L_{\rm CO}$, we have to be careful when selecting an appropriate stacking region.

\subsubsection{Stacking of Entire Galaxies}
\label{luminosity}

\begin{deluxetable}{l*{11}{c}}
\tablecolumns{12}
\tablecaption{Stacking of Entire Galaxy\label{t4}}
\tablehead{\colhead{Name} & \colhead{Area\tablenotemark{a}} & \colhead{$L_{\rm CO~2-1}$\tablenotemark{b}} & \colhead{SFR\tablenotemark{c}} \\
\colhead{} & \colhead{(kpc$^2$)} & \colhead{($10^6$~K~km~s$^{-1}~$pc$^2$)} & \colhead{($M_\odot$~yr$^{-1}$)}}
\startdata
M\,81 Dw\,A & 1.33 & $<$ 0.38 & 0.0004 \\
M\,81 Dw\,B & 1.79 & $<$ 0.64 & 0.0022 \\
DDO 053 & 1.75 & $<$ 0.37 & 0.0035 \\
DDO 154 & 7.74 & $<$ 1.00 & 0.0054 \\
DDO 165 & 11.47 & $<$ 1.46 & 0.0093 \\
HO I & 12.37 & $<$ 1.81 & 0.0093 \\
HO II & 26.69 & $<$ 2.83 & 0.0379 \\
IC 2574 & 83.85 & $<$ 8.20 & 0.0670 \\
NGC 2366 & 12.23 & $<$ 0.86 & 0.0532 \\
NGC 2403 & 39.40 & 26.79 $\pm$ 0.31 & 0.3131 \\
NGC 2976 & 19.21 & 13.89 $\pm$ 0.29 & 0.0908 \\
NGC 3077 & 7.78 & 3.54 $\pm$ 0.15 & 0.0689 \\
NGC 4214 & 14.75 & 3.21 $\pm$ 0.25 & 0.1041 \\
NGC 4236 & 59.54 & $<$ 3.53 & 0.1063 \\
NGC 4625 & 31.19 & 5.73 $\pm$ 0.65 & 0.0664 \\
NGC 5474 & 39.19 & $<$ 4.92 & 0.0683
\enddata
\tablenotetext{a}{Unprojected area sampled from this work.}
\tablenotetext{b}{$L_{\rm CO(2-1)}$ in sampled area from this work.}
\tablenotetext{c}{SFR(FUV+24) in sampled area from this work.}
\end{deluxetable}

We start with stacking the CO spectra over the entire (mapped) extent of each galaxy. To define the ``entire'' galaxy extent, we use the SFR distribution as guideline for the (most likely) distribution of molecular gas and CO emission. Unfortunately, this method does not provide definite sizes. FUV emission (the main tracer of SFR in dwarf galaxies outside massive star-forming regions) typically starts to flatten as function of galactocentric radius before reaching the background level. We therefore select for each galaxy a maximum galactocentric radius (typically between $1-2~R_{25}$) that includes most ($\sim95$\%) of the galaxy-integrated star formation. The so selected regions are highlighted by black contours in Figures \mbox{\ref{f1} \& \ref{f2}}.

We will show later that for some galaxies the as above selected region is not fully sampled by our CO map and may miss a significant fraction (in the worst cases $\sim 10-30$\%) of the total SFR as given in Table~\ref{t1}. This is especially true for NGC 5474 and NGC 2403, and to a lesser extent for Ho~II, NGC 2366, NGC 3077, NGC 4214, and NGC 4236. It is, however, not obvious how to correct for this effect. In the remainder of this paper we will therefore continue to refer to our measured $L_{\rm CO}$ as the total galaxy-integrated luminosity but urge the reader to keep in mind that the true value may be up to $\sim 30$\% higher for a (small) subset of our sample. The given uncertainties on $L_{\rm CO}$ include only the statistical uncertainties of fitting the stacked spectrum with Gaussian profiles. Uncertainties in the calibration (from instrumental and reduction methodology) may affect $L_{\rm CO}$ by up to 30\% \citep[see][]{Leroy2009a} and uncertainties in the distance will enter quadratically --- neither effect is included.

Figure~\ref{f3} shows the resulting spectra when stacking over the entire observed part of the galaxy (for the same galaxies shown in Figure~\ref{f2}). Table~\ref{t4} lists the (unprojected) area and the respective $L_{\rm CO}$ and SFR. Five galaxies, NGC 2403, NGC 2976, NGC 3077, NGC 4214, and NGC 4625 are robustly detected. These are the same galaxies that already showed emission for individual lines-of-sight (Section~\ref{sensitivity}). One galaxy, NGC 4236, may show a tentative signal which extends from $-8$ to $+20$ \kmpers, has peak intensity $\sim 2.4$ mK ($\sim 2.3\sigma$) over 2 channels, $I_{\rm CO~2-1} \approx 0.035$~\Kkmpers, and $L_{\rm CO,~2-1} \approx 2.1 \times 10^6$ \Kkmpers~pc$^2$, a factor $0.6$ below our quoted upper limit. This emission is not point-source-like because with a point source sensitivity of $L_{\rm CO} \sim 2.5 \times 10^4$ \Kkmperspc\ for this galaxy it would have been easily detected. All other galaxies remain undetected.

There are three galaxies, IC 2574, Ho~II, and NGC 5474, where we may have expected to find signal as these galaxies have properties similar to detected galaxies. The stacked spectrum of IC 2574 shows an enhancement peaking at $\sim 0-5$~\kmpers, with full width $\sim 16-18$ \kmpers, and peak intensity $\sim 4$ mK ($\sim 1\sigma$) for the 24\micron-selected regions or $\sim 1.3$ mK ($\sim 0.8\sigma$) over the entire galaxy. While the match between CO and \hi\ velocities is encouraging, the significance of this enhancement is too low to differentiate it from spurious emission. The stacked spectra of Ho~II and NGC 5474 show no signs of signal at a noise level of $1.9$ and $2.2$~mK per $2.6$ \kmpers\ channel, respectively.

\subsubsection{Stacking of 24\,$\mu$m-bright Regions}
\label{irregions}

\begin{deluxetable}{@{\extracolsep{-2pt}}l*{5}{c}}
\tablewidth{0.48\textwidth}
\tablecolumns{6}
\tablecaption{Stacking of 24$\mu$m-bright Regions\tablenotemark{a}\label{t5}}
\tablehead{\colhead{Name} & \colhead{Area} & \colhead{$L_{\rm 24\mu m}$\tablenotemark{b}} & \colhead{$L_{\rm CO~2-1}$} & \colhead{SFR} \\
\colhead{} & \colhead{\tiny{(kpc$^2$)}} & \colhead{\tiny{($10^6$~MJy~sr$^{\text{-}1}$~pc$^2$)}} & \colhead{\tiny{($10^6$~K~km~s$^{\text{-}1}~$pc$^2$)}} & \colhead{\tiny{($M_\odot$~yr$^{\text{-}1}$)}}}
\startdata
M\,81 Dw\,A & \nodata & \nodata & \nodata & \nodata \\
M\,81 Dw\,B & \nodata & \nodata & \nodata & \nodata \\
DDO 053 & 0.31 & 0.20 & $<$ 0.16 & 0.0016 \\
DDO 154 & \nodata & \nodata & \nodata & \nodata \\
DDO 165 & \nodata & \nodata & \nodata & \nodata \\
HO I & \nodata & \nodata & \nodata & \nodata \\
HO II & 1.82 & 1.35 & $<$ 0.38 & 0.0136 \\
IC 2574 & 2.39 & 1.54 & $<$ 0.52 & 0.0158 \\
NGC 2366 & 2.80 & 7.25 & $<$ 0.39 & 0.0403 \\
NGC 2403 & 33.01 & 50.87 & 26.04 $\pm$ 0.28 & 0.3021 \\
NGC 2976 & 7.63 & 17.78 & 12.10 $\pm$ 0.16 & 0.0856 \\
NGC 3077 & 6.13 & 19.10 & 3.53 $\pm$ 0.13 & 0.0679 \\
NGC 4214 & 5.99 & 15.73 & 2.33 $\pm$ 0.12 & 0.0934 \\
NGC 4236 & 7.15 & 5.75 & $<$ 1.01 & 0.0500 \\
NGC 4625 & 9.48 & 10.33 & 5.12 $\pm$ 0.31 & 0.0592 \\
NGC 5474 & 7.64 & 3.16 & $<$ 2.19 & 0.0375
\enddata
\tablenotetext{a}{This region includes all lines-of-sight with $I_{\rm 24\mu m} \geq 0.2$ MJy~sr$^{-1}$ at 13\arcsec\ resolution.}
\tablenotetext{b}{These units allow comparison to the trend $I_{\rm 24\mu m} \sim I_{\rm CO}$ found for massive spirals \citep{Schruba2011}.}
\end{deluxetable}

We make a final attempt to search for faint CO emission by stacking over regions that likely have the highest probability to contain molecular gas and may be bright in CO. These are regions rich in dust and showing signs of embedded high-mass star formation. We use the 24\micron\ intensity, $I_{\rm 24\mu m}$, as a tracer of these conditions and select all lines-of-sight that have $I_{\rm 24\mu m} \geq 0.2$ \MJypersr\ at 13\arcsec\ resolution. The adopted 24\micron\ level does not have a specific physical interpretation, but it is well ($\sim 4\sigma$) above the noise level of the 24\micron\ maps. IR emission tends to be faint in dwarf galaxies \citep[e.g.][]{Walter2007}. In the more massive dwarf galaxies of our sample this 24\micron\ level effectively separates the star-forming peaks from the rest of the galaxy. The smallest dwarf galaxies however do not reach this 24\micron\ intensity, and we omit them from this analysis. The thus selected regions are highlighted by red contours in Figure~\ref{f1}.

Table~\ref{t5} lists the results when stacking over these 24\micron-bright regions: the (unprojected) area, 24\micron\ and CO luminosity, and enclosed SFR. This method does not lead to new CO detections in addition to those galaxies already detected at individual lines-of-sight and over the entire galaxy. However, for non-detected galaxies it results in stronger upper limits on integrated quantities that scale with the size of the stacking regions, i.e., lower upper limits on $L_{\rm CO}$ and lower $L_{\rm CO} / \text{SFR}$ ratios.

\section{Scaling Relations for CO Luminosity}
\label{corelations}

\subsection{Comparison to Magellanic Clouds}
\label{individual}

\begin{deluxetable}{@{\extracolsep{-5pt}}lccl}
\tablewidth{0.48\textwidth}

\tablecolumns{4}
\tablecaption{Previous CO Observations of our Galaxy Sample\label{t6}}
\tablehead{\colhead{Name} & \colhead{Beam} & \colhead{$L_{\rm CO~1-0}$\tablenotemark{a}} & \colhead{Reference} \\
\colhead{} & \colhead{(arcsec)} & \colhead{($10^6\,$K$\,$km$\,$s$^{-1}\,$pc$^2$)} & \colhead{}}
\startdata
M\,81 Dw\,A & 45 & $<$ 0.16 & \citet{Young1995} \\
M\,81 Dw\,B & \nodata & \nodata & \nodata \\
DDO 053 & 55 & $<$ 0.94 & \citet{Leroy2005} \\
DDO 154 & 65 & $<$ 3.2 & \citet{Morris1978} \\
DDO 165 & 65 & $<$ 5.6 & \citet{Taylor1998} \\
Ho I & 65 & $<$ 3.8 & \citet{Taylor1998} \\
Ho II & 13$\times$65 & $<$ 9.9 & \citet{Elmegreen1980} \\
& 45 & $<$ 0.24 & \citet{Young1995} \\
& 55 & $<$ 1.2 & \citet{Leroy2005} \\
IC 2574 & 10$\times$65 & $<$ 10 & \citet{Elmegreen1980} \\
& 55 & 1.1 & \citet{Leroy2005} \\
NGC 2366 & 3$\times$65 & $<$ 1.9 & \citet{Elmegreen1980} \\
& 22 & $<$ 0.50 & \citet{Hunter1993} \\
& 22 & $<$ 0.04 & \citet{Albrecht2004} \\
& 55 & $<$ 1.3 & \citet{Leroy2005} \\
NGC 4214 & 45 & 0.38 & \citet{Young1995} \\
& 60 & 0.56 & \citet{Israel1997} \\
& 4$\times$55 & 0.73 & \citet{Taylor1998} \\
NGC 4236 & 11$\times$45 & $<$ 9.4 & \citet{Young1995} \\
NGC 4625 & 22 & 4.4 & \citet{Boker2003} \\
& 22 & 4.2 & \citet{Albrecht2004} \\
& 55 & 16 & \citet{Leroy2005} \\
NGC 5474 & 55 & $<$ 1.5 & \citet{Leroy2005}
\enddata
\tablenotetext{a}{Luminosities are given on main beam temperature scale ($T_{\rm mb}$) and are calculated assuming our adopted distances; line width and upper limit (3--4$\sigma$) on peak intensity are taken from the relevant reference.}
\end{deluxetable}

We begin with a comparison of our CO measurements (upper limits; Table~\ref{t4} \& \ref{t5}) to the Magellanic Clouds (Table~\ref{t2}). These are essentially the only low-metallicity systems that are well detected in CO over the full galaxy extent. We also list previous CO observations of our targets in Table~\ref{t6}. A direct comparison to our CO measurements is however not straightforward as previous observations covered only small fractions of the star-forming disk (often just a single pointing) and were strongly limited by sensitivity. The large scatter between individual literature measurements and compared to our values indicate that previous pointed observations have not been able to robustly constrain the galaxy-integrated CO luminosity of dwarf galaxies.

The galaxies that we detect in CO are comparable to (or exceed) the LMC in $M_{\rm HI}$, $M_{\rm B}$, SFR, and metallicity, but we are able to detect them at distances $D = 2.9 - 9.5$ Mpc. Galaxies that have not been detected when stacking over the entire galaxy extent, have $L_{\rm CO~2-1}$ upper limits $0.1 - 2.6$ times the CO luminosity of the LMC, $L_{\rm CO~1-0}^{\rm LMC} = 3.2 \times 10^6$ \Kkmperspc\ \citep{Fukui2008}. Our data is not sensitive enough to detect a CO luminosity comparable to the SMC, $L_{\rm CO~1-0}^{\rm SMC} = 1.6 \times 10^5$ \Kkmperspc\ \citep{Mizuno2006} if it is spread over many resolution elements. For the IR-selected regions, our CO sensitivity improved and is always sufficient to detect $L_{\rm CO~1-0}^{\rm LMC}$ and reaches down to $1-10$ times $L_{\rm CO~1-0}^{\rm SMC}$. For individual lines-of-sight we would have easily detected $L_{\rm CO~1-0}^{\rm SMC}$ but detect no such point sources for 11 of our 16 galaxies.

\subsection{Scaling Relations of $L_{\rm CO}$ with $M_{\rm B}$ and Metallicity}
\label{scalings}

\begin{figure}
\epsscale{1.06}
\plotone{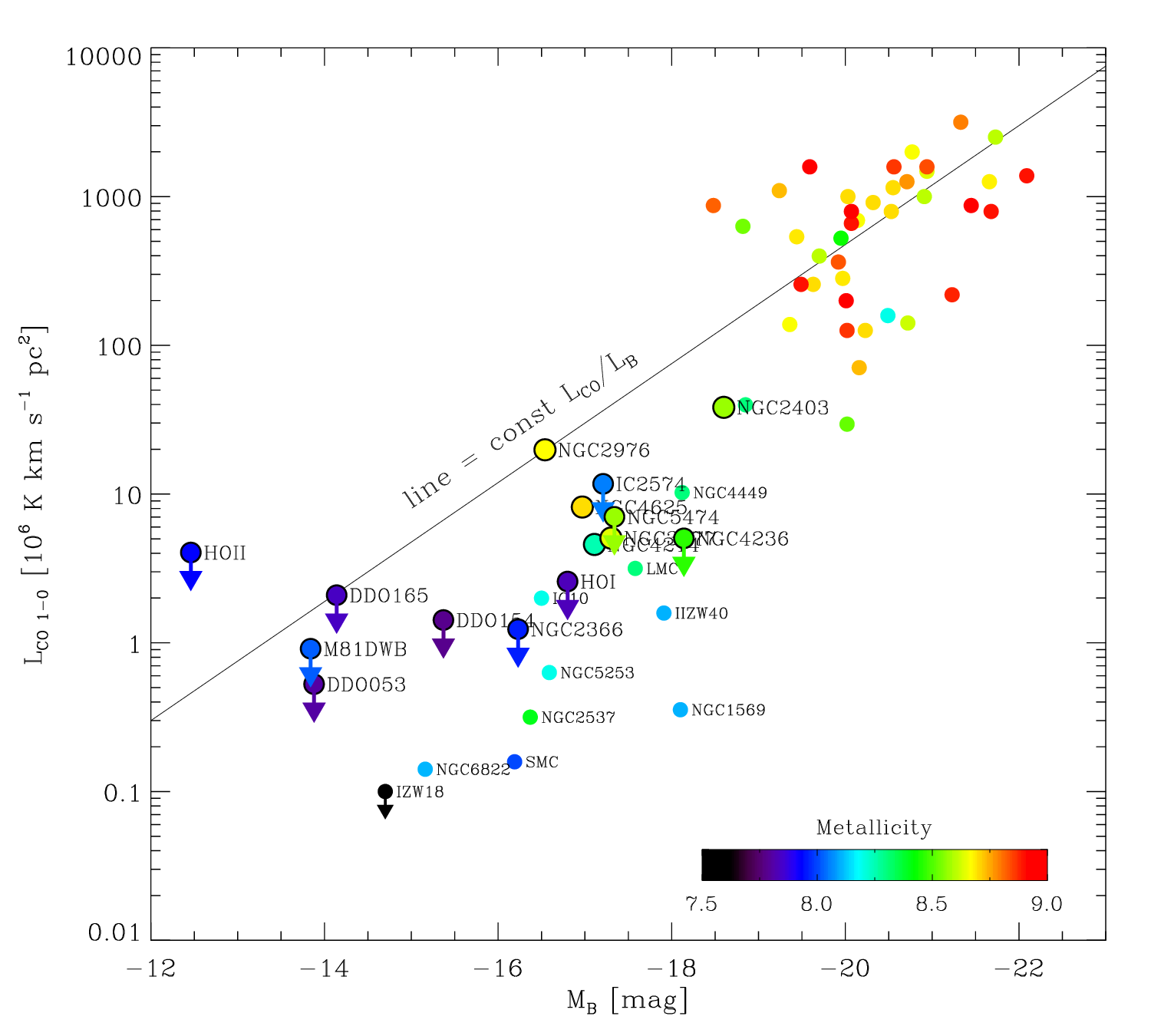}
\plotone{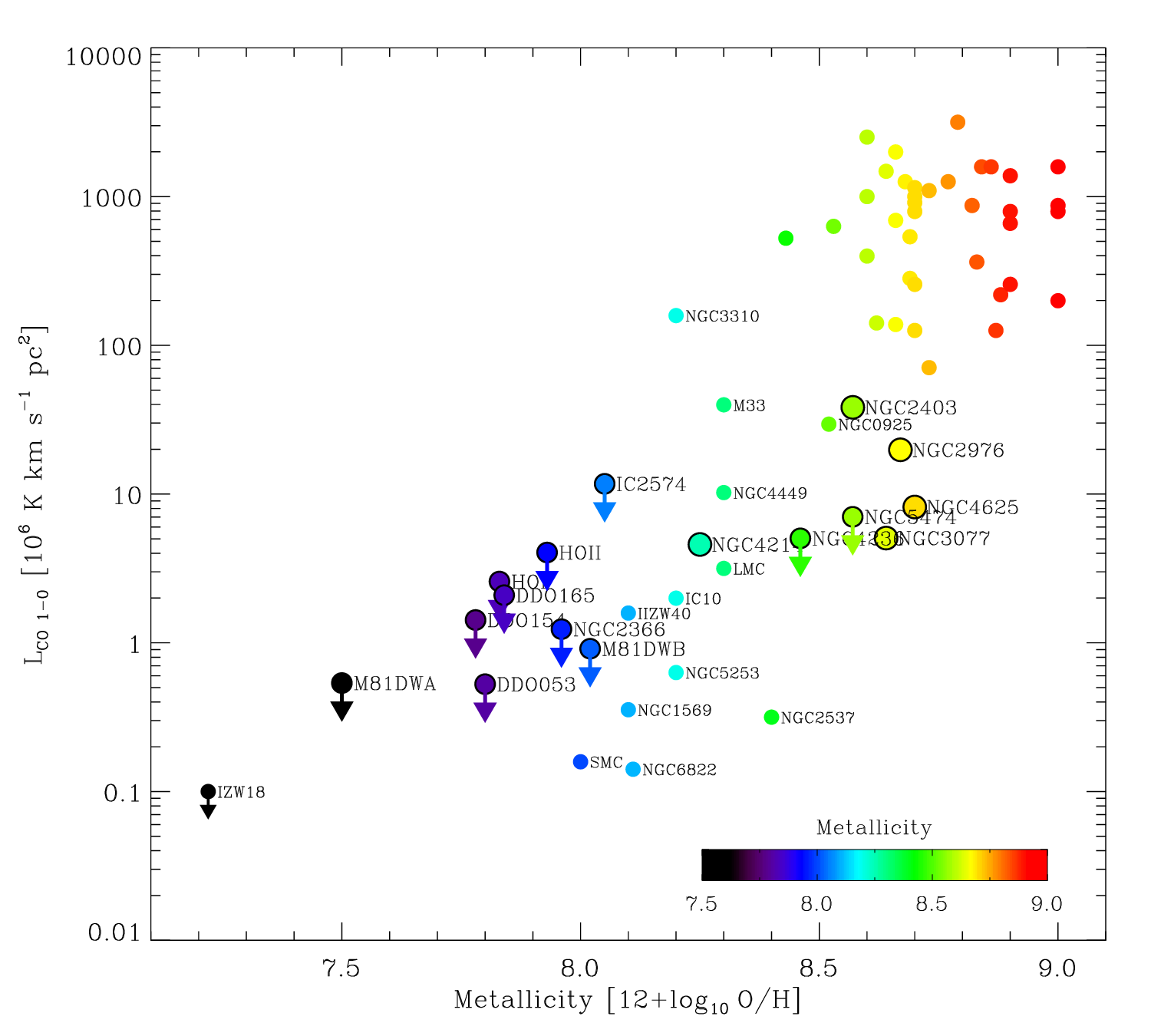}
\plotone{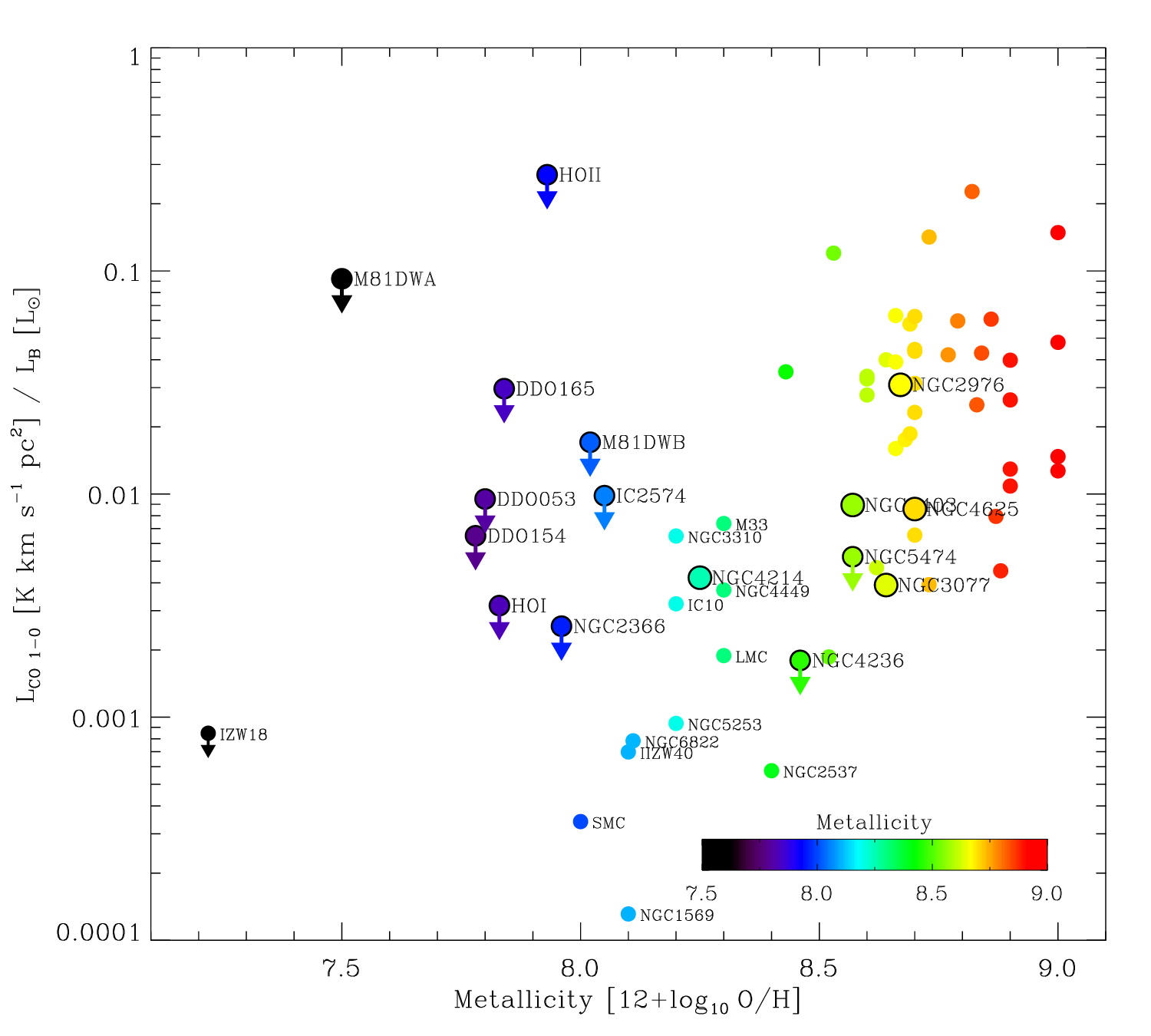}
\caption{{\em Scaling Relations for CO Luminosity.} Galaxy-integrated CO(1-0) luminosity as function of $B$-band magnitude (top) and metallicity (middle), and the ratio of $L_{\rm CO}$ and $L_{\rm B}$ as function of metallicity (bottom).
Bigger symbols show stacking results of this work, smaller symbols show a compilation of literature measurements. Color highlights metallicity. The solid line in the top panel shows a constant $L_{\rm CO} / L_{\rm B}$ ratio intersecting the bright galaxies. The trends are correlated by the well established luminosity-metallicity relation.\label{f4}}
\end{figure}

We use our robust estimates of the galaxy-integrated CO luminosity of dwarf galaxies to examine the relationship between $L_{\rm CO\,1-0}$, $B$-band magnitude, $M_{\rm B}$, and metallicity in Figure~\ref{f4}. In conjunction with our literature compilation our galaxy sample covers 5 orders of magnitude in $L_{\rm CO}$, 4 orders of magnitude in $L_{\rm B}$, $1.5$ orders of magnitude in metallicity, 5 orders of magnitude in star formation rate (SFR = $10^{-4} -10^1$ \Msunperyr), and 3 orders of magnitude in \hi\ mass ($M_{\rm HI} = 10^7 - 10^{10}$ \Msun).

The top panel of Figure~\ref{f4} shows $L_{\rm CO}$ as a function of $M_{\rm B}$. For guidance we show the solid line which highlights a constant scaling between $L_{\rm CO}$ and $L_{\rm B}$, set to intersect the bright galaxies. In the bright galaxies ($M_{\rm B} < -18$), $L_{\rm CO}$ and $M_{\rm B}$ track one another with a more-or-less fixed ratio, $\text{log}_{10}\ L_{\rm CO} [\Kkmperspc]\,/\,L_{\rm B} [L_\odot]= -1.5 \pm 0.4$, as one might expect for a simple scaling with galaxy mass \citep{Young1991, Leroy2005, Lisenfeld2011}. The dwarf galaxies ($M_{\rm B} \geq -18$) on the other hand lie below the solid line. They are ``underluminous'' in CO, i.e., their ratios are systematically smaller, $-2.7 \pm 0.6$, than those of massive galaxies. Despite this trend, $L_{\rm CO}$ and $M_{\rm B}$ are strongly correlated with (absolute) rank correlation coefficient $r_{\rm corr} = 0.79$. 

The middle panel of Figure~\ref{f4} shows $L_{\rm CO}$ as function of metallicity. There is a dramatic drop in $L_{\rm CO}$ by $3-4$ orders of magnitude over a small range of metallicities. This drop is to first order caused by the much smaller mass and size of dwarf galaxies. However, due to the strong luminosity--metallicity relation for dwarf irregulars \citep[e.g.,][]{Lee2006, Guseva2009}, it is also correlated to $M_{\rm B}$ (i.e., the top panel). The rank correlation coefficient is $r_{\rm corr} = 0.60$.

The bottom panel of Figure~\ref{f4} shows the ratio $L_{\rm CO} / L_{\rm B}$ as function of metallicity. Plotting $L_{\rm CO} / L_{\rm B}$ should remove most of the mass and size dependence seen in the above panels. The decreasing trend of $L_{\rm CO} / L_{\rm B}$ with deceasing metallicity clearly shows that dwarf galaxies are also ``underluminous'' in CO in a normalized sense. 

From the study of a large sample of literature CO data, \citet{Taylor1998} suggested a ``detection threshold'' for CO below $\text{12+log\,O/H} \approx 8.0$, about the metallicity of the SMC. Our data do not overcome this threshold: All of our galaxies with lower metallicity remain undetected. However, given the decreasing trend of $L_{\rm CO} / L_{\rm B}$ with decreasing metallicity and the fact that our data is not sensitive enough to detect a galaxy like the SMC at a distance of $D = 4$ Mpc leaves open the question whether the proposed threshold is of observational or physical origin.

\subsection{CO and Tracers of Star Formation}
\label{sflaw}

\begin{figure*}
\epsscale{1.0} \plotone{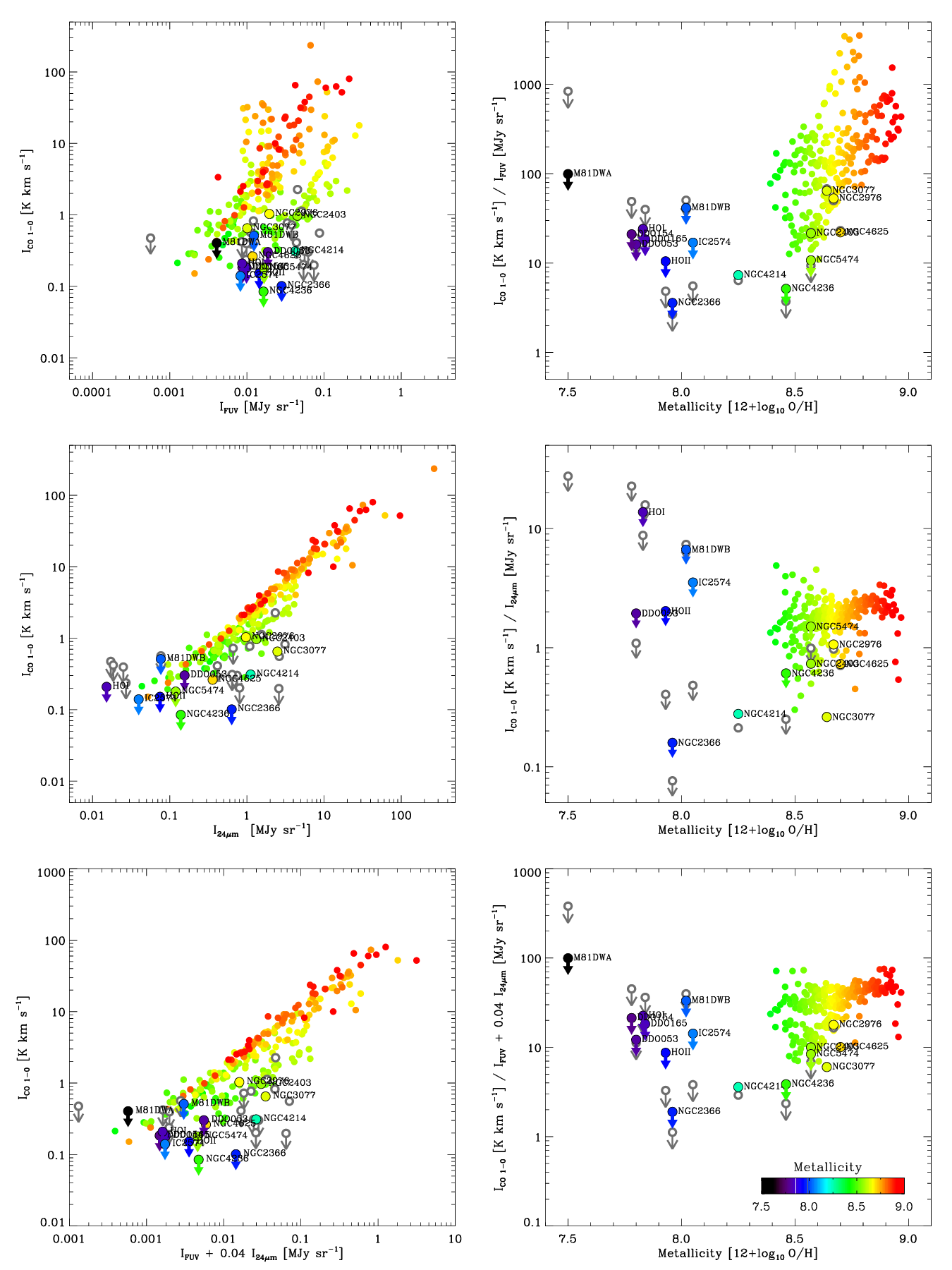}
\caption{{\em CO Emission and Tracers of Star Formation.} Left column: CO(1-0) intensity as function of FUV and 24\micron\ intensity, and a combination of FUV and 24\micron\ used to determine the SFR. Right column: Intensity ratios of CO and FUV or 24\micron\ as function of metallicity. Bigger symbols show stacking results for dwarf galaxies derived over the entire galaxy extent (colored symbols) or over 24\micron-bright regions (gray symbols). Smaller symbols show azimuthally averaged radial profiles in massive spiral galaxies. CO emission is well correlated with emission of SFR tracers, especially with 24\micron. The ratios CO/FUV and CO/24\micron\ show systematically smaller values in low-metallicity environments as is typical for dwarf galaxies.\label{f5}}
\end{figure*}

Figure~\ref{f5} shows the relationship between CO emission and tracers of recent star formation, FUV and 24\micron, together with a combination of FUV and 24\micron\ often used to estimate the SFR (see Section~\ref{sftracers}). Panels in the left column show the correlations between observables $I_{\rm CO}$, $I_{\rm FUV}$, and $I_{\rm 24\mu m}$, panels in the right column show the ratios of $I_{\rm CO}$ and $I_{\rm FUV}$ or $I_{\rm 24\mu m}$ as function of metallicity. For our dwarf galaxies (bigger symbols) we show both the stacking results derived over the entire galaxy (colored by metallicity) as well as the values derived from stacking over the 24\micron-bright regions (gray symbols). For comparison we also show azimuthally averaged radial intensity profiles for spiral galaxies (smaller colored symbols) which are taken from \citet{Schruba2011}. The plots shown here are similar to plots of  the ``star formation law'', i.e., plots of $\Sigma_{\rm SFR}$ versus $\Sigma_{\rm H2}$, though these are typically presented with axes interchanged and show data that are corrected for inclination (which we have not done here).

One of the results from \citet{Schruba2011} was that spiral galaxies exhibit a strong correlation between $I_{\rm CO}$ and $I_{\rm 24\mu m}$ (see middle panel in left column). This has been interpreted as a direct link between molecular gas as traced by CO emission and SFR, which is mostly deeply embedded and traced by 24\micron\ emission. The ratio of CO/24\micron\ is roughly constant inside galaxies and shows only little variation between galaxies and as function of metallicity where CO is detected (middle panels). The ratio of CO/FUV also shows little scatter inside individual galaxies but can vary  significantly between galaxies (upper panels). The large scatter in CO/FUV and the strong scaling with metallicity reflects the strong susceptibility of FUV emission to dust attenuation. In low metallicity environments CO/FUV is low because CO abundance and thus $I_{\rm CO}$ is low and at the same time, low dust abundance and low attenuation cause $I_{\rm FUV}$ to be relatively high. The ratio of $I_{\rm CO}$/($I_{\rm FUV}$+0.04$I_{\rm 24\mu m}$) which is proportional to the H$_2$ depletion time, $\tau_{\rm dep}$, is to first order constant $\sim 1.8$~Gyr for environments with \mbox{12+log$_{10}$\,O/H} $\approx 8.7$ and shows little dependence on metallicity in environments with about solar metallicity \citep[see also][]{Bigiel2008, Bigiel2011, Leroy2011}.

Our dwarf galaxies however do show deviations from the trends observed for radial profiles of more massive galaxies. The CO, FUV, and 24\micron\ intensities measured in the dwarf galaxies are close to the lowest intensities measured in the radial annuli of more massive galaxies. In addition, the ratios of CO/FUV and CO/24\micron\ are shifted to smaller values. For our detected galaxies the ratios are a factor $5-10$ below the ratios found in more massive galaxies. The data of our undetected galaxies are scattered but for galaxies with sensitive CO upper limits they are also shifted toward low CO/FUV and CO/24\micron\ ratios. Dwarf galaxies exhibit enhanced signatures of star formation (both embedded and unobscured) per unit CO brightness as compared to large star-forming galaxies.

\section{Implications for CO-to-H$_2$ Conversion Factor}
\label{xcofactor}

A serious complication in studying the molecular content of dwarf galaxies arises in how to relate the observed CO luminosities to H$_2$ masses. Applying a Galactic CO-to-H$_2$ conversion factor, $\alpha_{\rm CO,\,Gal}$, to dwarf galaxies that have been detected in CO results in low H$_2$ masses \citep{Taylor1998, Mizuno2001, Leroy2007}. The resulting H$_2$ masses are so low that to explain the observed SFRs the conversion of H$_2$ to stars would need to be on average $10-100$ times more efficient than in Galactic environments --- a condition that seems unlikely \citep[e.g.,][]{Bolatto2011}.

The detection of excess ionized carbon and infrared to millimeter dust emission around star-forming regions \citep{Madden1997, Pak1998, Rubin2009, Cormier2010, Israel2011} indicates that CO may not trace all H$_2$ at low metallicity \citep{Maloney1988, Israel1997, Bolatto1999, Wolfire2010}. Because H$_2$ can self-shield, its abundance is basically a function of its formation time (which depends on metallicity), however, CO cannot self-shield and exists only in regions that are sufficiently shielded by dust from the interstellar radiation field \citep{Glover2010, Glover2011}. $\alpha_{\rm CO}$ is therefore assumed to be a strong function of metallicity and radiation field strength, although robust functional parametrizations of these dependences are still lacking (but see \citealp{Shetty2011a, Shetty2011b} and \citealp{Narayanan2011} for recent theoretical works).

In the following we will discuss three different methods that have been applied to estimate $\alpha_{\rm CO}$ in external galaxies. In particular we are interested in the metallicity dependence of $\alpha_{\rm CO}$. Whenever possible, we parametrize this dependence by a linear function between $\text{log}_{10}\,\alpha_{\rm CO}$ and $(\text{12+log\,O/H})$, i.e.,

\begin{eqnarray}
\text{log}_{10}\,\alpha_{\rm CO} = \text{log}_{10}\,A + N \times (12+\text{log}_{10}\,\text{O/H} - 8.7)
\label{e1}
\end{eqnarray}

\noindent where the normalization $A$ gives $\alpha_{\rm CO}$ at $\text{12+log\,O/H} = 8.7$ and $N$ is the slope.

Before we begin, we have to caution the reader that gas phase metallicities bear considerable uncertainties. Different empirical and theoretical calibrations can result in systematic discrepancies in estimated metallicities as large as $0.1-0.7$ dex \citep{Kewley2008}. However, once a specific calibration is selected the relative ordering of individual galaxies and derived slopes are more robust. For our new data and our literature compilation we have tried to maximize homogeneity of metallicity estimates (i.e., we use the metallicity calibration described in Section~\ref{metallicities} whenever possible). To be on the save side, we thus concentrate our discussion on the slope of the parametrization between $\alpha_{\rm CO}$ and metallicity. For the same reason a direct comparison of our results to the literature is hindered as previous studies applied a variety of different metallicity calibrations that may even vary within individual studies.

\subsection{Different Methods to Estimate the CO-to-H$_2$ Conversion Factor}

\subsubsection{Virial Method}

The classic method to derive $\alpha_{\rm CO}$ uses high resolution CO observations that are capable of resolving individual molecular clouds \citep[e.g.,][]{Solomon1987}. Under the assumption that a CO-bright core is in virial equilibrium, its observed linewidth and size can be converted into a virial mass, $M_{\rm vir}$, and from that $\alpha_{\rm CO} \equiv M_{\rm vir} / L_{\rm CO}$. Early work by \citet{Wilson1995}, \citet{Arimoto1996}, and \citet{Boselli2002} have applied this method to a handful of Local Group galaxies and found a weak metallicity dependence of $\alpha_{\rm CO}$ with slopes flatter than $-1$. This metallicity dependence however has not been confirmed by the recent studies of \citet{Blitz2007} and \citet{Bolatto2008}. They re-analyze a large set of literature data aiming at maximizing homogeneity of their analysis and carefully correcting for finite spatial and spectral resolution. They derive a distribution of $\alpha_{\rm CO}$ values that scatters without systematic trend around $0.5 -5~\alpha_{\rm CO,\,Gal}$ (the green striped region in Figure~\ref{f7}; we describe this figure later in detail). It has to be emphasized that the virial method likely leads to a significant underestimate of the total H$_2$ mass of GMCs in low-metallicity environments. This bases on the idea that with decreasing metallicity the CO-bright cores shrink to (much) smaller size than the surrounding H$_2$ clouds \citep[e.g.,][]{Bolatto1999, Wolfire2010, Shetty2011a, Shetty2011b}.

\begin{figure*}
\epsscale{0.7} \plotone{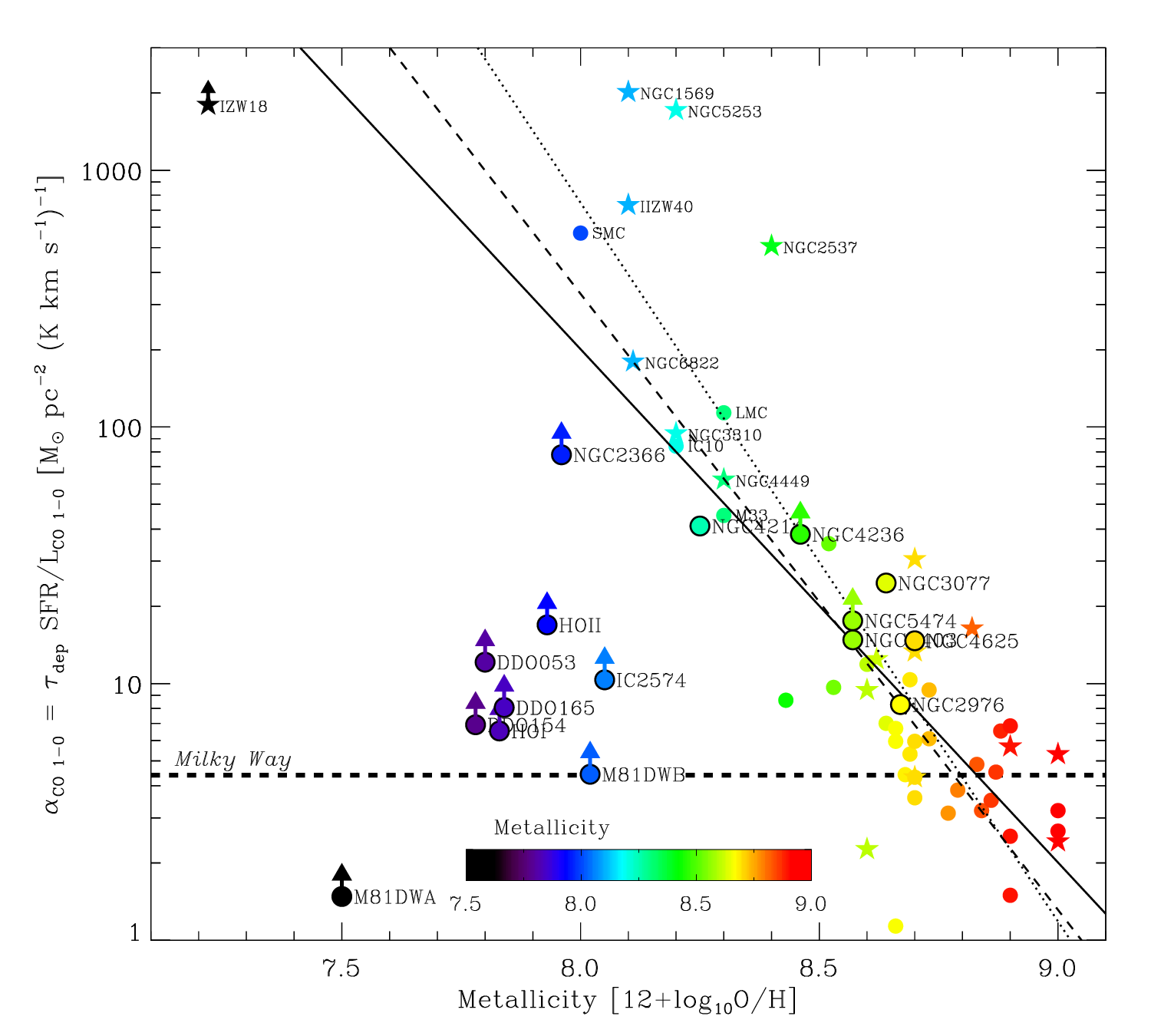}
\caption{{\em Metallicity Dependence of the CO-to-H$_2$ Conversion Factor.} $\alpha_{\rm CO}$ is derived from the ratio of observed SFR scaled by a constant H$_2$ depletion time, $\tau_{\rm dep}= 1.8$ Gyr, and the observed CO luminosity, $L_{\rm CO}$. Bigger symbols show galaxy-integrated measurements of dwarf galaxies from this work, smaller symbols show data for our literature compilation with starbursts highlighted by stars. The horizontal dashed line shows the Galactic conversion factor and the diagonal lines show regression fits: to all galaxies (dotted line), to all non-starburst galaxies (dashed line), and exclusively to the HERACLES sample (solid line).\label{f6}}
\end{figure*}

\subsubsection{Dust Modeling}

This method uses IR observations and dust modeling to estimate the gas mass and distribution which has the advantage that it is independent of CO emission \citep{Thronson1988, Israel1997, Dame2001, Leroy2007, Leroy2009a, Leroy2011, Gratier2010b, Bolatto2011}. It builds on the assumption that gas and dust are well mixed and $\alpha_{\rm CO}$ is derived from $M_{\rm dust} \equiv \text{DGR} \times (M_{\rm HI} + \alpha_{\rm CO} L_{\rm CO})$, where DGR is the dust-to-gas ratio. By modeling the dust distribution the local H$_2$ mass can be inferred from $M_{\rm dust}$ (after subtraction of local \hi) by either fixing the DGR in quiescent, non-star-forming regions (assumed to be H$_2$-free) or by simultaneous optimizing $\alpha_{\rm CO}$ and DGR such that the scatter between $M_{\rm dust}$/DGR and $M_{\rm HI} + \alpha_{\rm CO} L_{\rm CO}$ gets minimized. Early work by \citet{Israel1997} implied a strong metallicity dependence of $\alpha_{\rm CO}$ with slope of $-2.7 \pm 0.3$. Recent work by \citet{Gratier2010b}, \citet{Leroy2011}, and \citet{Bolatto2011} did confirm a strong increase of $\alpha_{\rm CO}$ at low metallicities, although their results vary in absolute terms, proposed functional form, and are systematically smaller than the $\alpha_{\rm CO}$ values derived by \citet{Israel1997}. The $\alpha_{\rm CO}$ values derived by these studies lie within the blue striped region in Figure~\ref{f7}. The lowest metallicity galaxy to which this method has been applied is the SMC, for which large amounts of H$_2$ have been inferred implying $\alpha_{\rm CO}$ values $10-100$ times the Galactic value \citep{Leroy2011, Bolatto2011}.

Disadvantages of this method are that it is susceptible to variations of the FIR emissivity of dust grains and variations in DGR between dense, star-forming regions and low density, quiescent regions. There are indications that the emissivity is enhanced in dense regions \citep[e.g.][]{Paradis2009, Planck2011} which would cause an overprediction of $\alpha_{\rm CO}$ on scales of individual star-forming regions.  Second, because dust enrichment of the ISM by stars seems insufficient to explain observed dust abundances, it is proposed that most dust forms in the ISM, presumably in the densest regions \citep{Dwek1998, Draine2007}. If this dust is only slowly transported into the lower density ISM then this would also lead to an overprediction of $\alpha_{\rm CO}$. The need for sensitive, matched high resolution data to make the analysis robust limits this method to nearby galaxies and makes observations time consuming.

\subsubsection{Constant SFE}

An alternative method to constrain the H$_2$ mass is to assume that the conversion of H$_2$ to stars is independent of environment, i.e., assuming a constant H$_2$ depletion time, $\tau_{\rm dep}$, or a constant star formation efficiency (SFE; the inverse of $\tau_{\rm dep}$).  $\alpha_{\rm CO}$ is then given by $\alpha_{\rm CO} \equiv \tau_{\rm dep} \times \text{SFR} / L_{\rm CO}$. This approach has been applied even when it was still considered very uncertain how to relate CO to H$_2$ in our Galaxy \citep{Rana1986}. We will apply this method to our sample of nearby galaxies in the remainder of this paper. The idea is encouraged by several observations: (a) the accumulating evidence that star formation in molecular clouds is independent from environment as indicated by the similarity of molecular cloud properties in our and nearby galaxies \citep{Blitz2007, Bolatto2008, Fukui2010}, the universality of SFE per free-fall time in clouds of different mass and density \citep{Krumholz2007}, and evidence in favor of a universal initial stellar mass function \citep{Bastian2010}; and (b) the remarkably constant scaling between H$_2$ and SFR on $\sim$~kpc scales observed in a large set of nearby spiral galaxies \citep[e.g.,][]{Bigiel2008, Bigiel2011, Leroy2008, Schruba2011}. A drawback of this method is that it requires that the correlation between H$_2$ and SFR established in spiral galaxies continues to hold in dwarf galaxies. This makes the method less rigorous than direct attempts to trace H$_2$ but also makes it available for a much larger sample of galaxies including distant galaxies.

Currently we are not able to conclude that $\tau_{\rm dep}$ is truly constant. Observations of strongly variable star formation histories and starbursts readily indicate that it does not hold in all environments \citep[e.g.][]{Lee2009a, Weisz2011}. However, recent theoretical considerations by \citet{Krumholz2011} and \citet{Glover2012} provide a motivation in favor of a constant H$_2$/SFR ratio. Although some of these methods do question if H$_2$ is fundamental for star formation, they also argue that H$_2$ will be a good tracer of star-forming regions. This is because the \hi\ to H$_2$ transition and the drop in gas temperature which makes clouds susceptible to gravitational instabilities occur under similar conditions that are to first order set by dust shielding of the interstellar radiation field.

\subsection{New \& Literature Measurements}

\begin{figure*}
\epsscale{0.7} \plotone{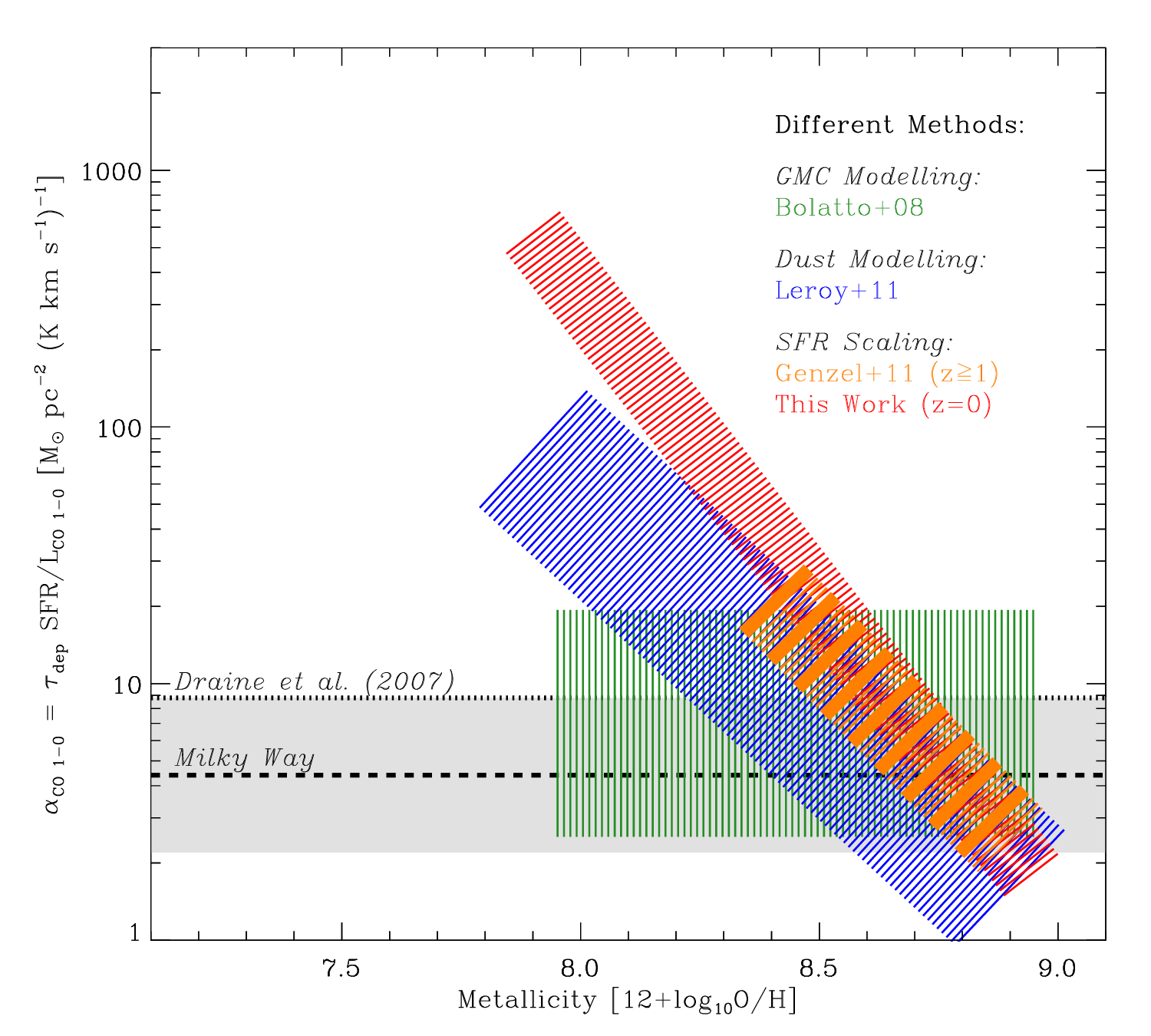}
\caption{{\em Trends of the Metallicity Dependence of the CO-to-H$_2$ conversion factor $\alpha_{\rm CO}$.} Striped bands indicate the range of $\alpha_{\rm CO}$ values derived from employing different methods (see text). The width in the bands indicate roughly the scatter of individual measurements.
\label{f7}}
\end{figure*}

In the following we explore the implications for CO if $\tau_{\rm dep}$ is indeed constant. The average value for spiral galaxies in the HERACLES sample with about solar metallicity varies in the range of  $\sim 1.8$ to $2.35$ Gyr \citep{Bigiel2008, Bigiel2011, Leroy2008, Schruba2011}. This natural variation is due to minor differences from galaxy to galaxy, and depends on exactly which lines-of-sight are included, and in particular the weighting method employed (i.e., galaxy average, radial rings, or pixel averages). We adopt here a value of $\tau_{\rm dep} = 1.8$ Gyr which is derived on the basis of entire galaxy averages. Figure~\ref{f6} shows the resulting $\alpha_{\rm CO}$ values as function of metallicity. In this plot we show galaxy-integrated values, bigger symbols show our measurements of dwarf galaxies (Table~\ref{t4}) and smaller symbols show data from our literature compilation (Table~\ref{t2}). Color coding highlights metallicity as in previous plots. Star symbols indicate galaxies that are labeled in the literature as starbursts.

The derived $\alpha_{\rm CO}$ values strongly depend on metallicity. For galaxies with \mbox{12+log$_{10}$\,O/H} $\gtrsim 8.6$, we find $\alpha_{\rm CO} \sim \alpha_{\rm CO,\,Gal}$ although with $\sim 0.3$ dex (factor 2) scatter. For galaxies with lower metallicity, $\alpha_{\rm CO}$ increases strongly with decreasing metallicity. For dwarf galaxies with \mbox{12+log$_{10}$\,O/H} $\lesssim 8.6$, even though most of them remain undetected in CO, we can readily exclude $\alpha_{\rm CO} \sim \alpha_{\rm CO,\,Gal}$. The few dwarf galaxies with CO detection suggest $\alpha_{\rm CO} \gtrsim 10~\alpha_{\rm CO,\,Gal}$ at \mbox{12+log$_{10}$\,O/H} $\lesssim 8.4$. We emphasize that the derived $\alpha_{\rm CO}$ values for dwarf galaxies with $Z / Z_\odot \sim 1/2 - 1/10$ are $1-2$ orders of magnitude higher than $\alpha_{\rm CO}$ values derived for massive spirals with $Z / Z_\odot \sim 1$.

\begin{deluxetable}{lccc}
\tablecolumns{4}
\tablecaption{Metallicity Dependence of $\alpha_{\rm CO}$ assuming a constant SFE\tablenotemark{a}\label{t7}}
\tablehead{\colhead{Selected Data} & \colhead{Value at} & \colhead{Slope of} & \colhead{Scatter} \\
\colhead{} & \colhead{12+logO/H=8.7} & \colhead{Regression} & \colhead{(dex)}}
\startdata
complete sample & & \\
$\bullet$\hspace{2mm}all galaxies & $ 8.2 \pm  1.0$ & $-2.8 \pm  0.2$ & 0.13 \\
$\bullet$\hspace{2mm}non-starbursts & $ 6.9 \pm  1.0$ & $-2.4 \pm  0.3$ & 0.10 \\[2mm]
HERACLES sample & & \\
$\bullet$\hspace{2mm}all galaxies & $ 8.0 \pm  1.3$ & $-2.0 \pm  0.4$ & 0.10 \\
$\bullet$\hspace{2mm}non-starbursts & $ 7.1 \pm  1.2$ & $-2.0 \pm  0.4$ & 0.09
\enddata
\tablenotetext{a}{Using bisecting linear regression of $\text{log}_{10}\,\alpha_{\rm CO} = \text{log}_{10}\,A + N \times (12+\text{log}_{10}\,\text{O/H}-8.7)$; see Eq.~(\ref{e1}). Uncertainties are determined by repeatedly adding random noise of $0.1$~dex to $12+\text{log}_{10}\,\text{O/H}$ ($x$-axis) and $0.3$~dex to $\text{log}_{10}\,\alpha_{\rm CO}$ ($y$-axis).}
\end{deluxetable}

We attempt to parametrize this dependence by fitting function of the form given in Eq.~(\ref{e1}). We use a bisecting linear regression to determine the best-fitting parameters. Uncertainties are determined from a Monte Carlo analysis. We have repeatedly added Gaussian noise to $\alpha_{\rm CO}$ with log-normal standard deviation of $0.3$~dex and to $\text{12+log}_{10}\,\text{O/H}$ with standard deviation of $0.1$~dex and re-fitted the perturbed data. The quoted uncertainties correspond to the standard deviation of 100 such derived best-fit parameters. Table~\ref{t7} lists the resulting normalizations and slopes together with the scatter of the data orthogonal to the best-fit regression. We divide our galaxy sample in two groups: ``starbursts'' and ``non-starbursts'', and evaluate the HERACLES galaxies and the complete sample separately. This may help to minimize biases due to inhomogeneous data sets. We expect the smallest systematics for the HERACLES sample including only non-starburst galaxies. We separate the starbursts because they likely violate our assumption of a constant $\tau_{\rm dep}$ having SFR in excess of their H$_2$ content.

The best-fit regressions depend somewhat on the particular galaxy sample; see Figures~\ref{f6} and Table~\ref{t7}. For this analysis we neglect the upper limit measurements as they are not stringent enough to affect our best-fits. For the HERACLES sample we determine a slope of $-2.0 \pm 0.4$ roughly independent whether startbursts are included or not (solid line), but with larger uncertainties due to the relative small dynamic range sampled by the detected galaxies. For the complete galaxy sample, the slope is steeper. We determine a slope of $-2.4 \pm 0.2$ for the non-starbursts (dashed line) and $-2.8 \pm 0.2$ for all galaxies (dotted line). We consider the latter result uncertain and potentially biased high because it is driven by a handful galaxies that are currently undergoing a starburst and have CO measurements from \citet{Taylor1998}, measurements that have been made prior to the latest generation of sensitive millimeter receivers. The scatter of the data to the best-fit relations is $0.09 - 0.12$ dex ($\sim 30$\%) which is significant smaller than the scatter of $\sim 0.3$ dex in the ratio $\alpha_{\rm CO} / \alpha_{\rm CO,\,Gal}$ for galaxies with \mbox{12+log$_{10}$\,O/H} $\gtrsim 8.6$. In this sense a steep increase of $\alpha_{\rm CO}$ with decreasing metallicity is much favored as compared to a constant value. The trend fitted to the ``complete, non-starburst'' sample and its associated uncertainty is indicated as red striped region in Figure~\ref{f7}.

The recent study by \citet{Genzel2012} also applied the assumption of a constant SFE. They analyzed a sample of star-forming galaxies at redshift $z \sim 1-2$ and determined a slope of $-1.9 \pm 0.67$; the orange striped region\footnote{Note that we changed the normalizations of the \citeauthor{Genzel2012} parameterizations to match the data plotted in their Figure~3. For the high redshift sample we increased the normalization by $0.3$ dex, for the combined sample we decreased it by $0.07$ dex.} in Figure~\ref{f7}. They also combined their distant galaxy sample with the dust-inferred $\alpha_{\rm CO}$ measurements from \citet{Leroy2011} which reduces their slope  to $-1.3 \pm 0.25$. The decrease in slope is basically driven by two galaxies, M31 and the SMC, and their result may be affected by combining two different methods. \\

\subsection{Comparison}

Approximate trends for the metallicity dependence of $\alpha_{\rm CO}$ derived from the three discussed methods and their intrinsic scatter are indicated in Figure~\ref{f7}. At solar metallicities the three methods give roughly consistent results within their uncertainties. Toward lower metallicities the three methods however predict different trends for the dependence of $\alpha_{\rm CO}$ on metallicity. The $\alpha_{\rm CO}$ values derived from the virial method (green striped region) show no systematic trend with metallicity and exhibit roughly an order of magnitude scatter. The dust-inferred $\alpha_{\rm CO}$ values (blue striped region) and the $\alpha_{\rm CO}$ values derived from scaling the SFR (red and orange striped regions) however do show a strong systematic increase toward low metallicities. For dwarf galaxies with $Z / Z_\odot \sim 1/2 - 1/10$ the dust-inferred $\alpha_{\rm CO}$ values are roughly a factor of 10 larger than the Galactic value but with $\sim 1$ order magnitude scatter, for the SFR-scaled $\alpha_{\rm CO}$ values they are a factor $10-100$ larger.

In addition to changes in the distribution of CO emitting gas at low metallicities, changes in the physical conditions may also affect the excitation temperature and thus the (relative) brightness of the \coone\ and \cotwo\ transitions. For molecular clouds in the Magellanic Clouds, line ratios of $R_{21} \sim 1.0 - 1.5$ are frequently observed \citep{Bolatto2000, Bolatto2003, Israel2003, Israel2005}. This is typically (but not always) ascribed to higher fractions of warm and optically thin gas as CO is less shielded against the interstellar radiation field. In diffuse gas, values of $R_{21}$ can be as high as $\sim 3$, however, these diffuse regions are typically faint and contribute only a minor fraction to the total CO emission \citep[e.g.,][]{Israel2003}. Thus our choice of using a single value of $R_{21} = 0.7$ when converting the HERACLES \cotwo\ line intensities to \coone\ values may bias our results to over-predict the true \coone\ line intensity in the HERACLES low metallicity galaxies by a factor $\lesssim 2$. The lack of a systematic offset to the sample of literature galaxies in Figures~\ref{f6} that all use \coone\ data suggest that variations in $R_{21}$ do not dominate our analysis.

The discrepancy between $\alpha_{\rm CO}$ values derived from the viral method and those derived from dust or SFR at low metallicities can be explained by considering the spatial scale which these methods operate on. The virial method operates on the small spatial scales of CO-bright cores of molecular clouds. At low metallicity these CO cores shrink as CO in low density gas gets dissociated while H$_2$ can survive there via self-shielding. Applying such $\alpha_{\rm CO}$ values to the total CO luminosity of a galaxy thus traces only the H$_2$ mass within the high density CO-bright cores and at low metallicity will inevitably fail to trace the total H$_2$ mass. On the other hand, the $\alpha_{\rm CO}$ values derived from dust or SFR are sensitive to CO-dark H$_2$ and can trace H$_2$ on spatial scales of the size of star-forming regions and larger.

But why are the $\alpha_{\rm CO}$ values derived from dust and SFR different? We infer $\alpha_{\rm CO}$ under the assumption of a fixed H$_2$/SFR ratio, i.e., $\tau_{\rm dep} = 1.8$ Gyr, that spans from massive spirals to low-mass, low-metallicity dwarfs. If this assumption breaks down then we will mis-attribute variations in H$_2$/SFR to variations in $\alpha_{\rm CO}$. In models in which the star formation efficiency is set by the free-fall time, metallicity can affect H$_2$/SFR. This is pointed out by \citet{Gnedin2009}, \citet{Gnedin2011}, and \citet{Feldmann2011b}; they show that the gas densities containing H$_2$ vary strongly with metallicity and radiation field, and thus free-fall times of clouds containing H$_2$ are not constant. While this can cause significant scatter in H$_2$/SFR on cloud scales, on large ($\sim$ kpc) scales variations are expected to be much smaller. In environments with metallicity $Z / Z_\odot \sim 1/10$ and radiation field $U / U_\odot \sim 10 - 100$, free-fall times and thus H$_2$/SFR are reduced by (only) a factor $2-3$. This is consistent with \citet{Krumholz2011}; they find $\Sigma_{\rm SFR} / \Sigma_{\rm H2}$ to be constant within a factor~2 for ranges of $\Sigma_{\rm H2} = 0.1 - 100$ \Msunperpc\ and $Z / Z_\odot = 1 - 1/10$.

In the handful of studies that attempt to account for $\alpha_{\rm CO}$ variations and measure H$_2$/SFR in low-metallicity galaxies \citep[e.g.,][]{Gratier2010a, Gratier2010b, Bolatto2011}, there are suggestions that H$_2$/SFR is up to $2-5$ times lower in local dwarfs at $Z / Z_\odot \sim 1/5$. A factor of $\sim 2-5$ adjustment will not perfectly reconcile the various $\alpha_{\rm CO}$ measurements at the lowest metallicities but can provide rough agreement at $Z / Z_\odot \sim 1/2 - 1/5$. In this case Figure~\ref{f7} and similar plots combine two important trends: variations in $\alpha_{\rm CO}$ and in $\tau_{\rm dep}$. More detailed work comparing the SFR to H$_2$ estimated via independent tracers like dust, \mbox{\rm [\ion{C}{2}]}, or gamma rays will be needed to refine this approach.

As noted above, the relative alignment of our galaxy sample along the $x$-axis (metallicity) in Figure~\ref{f7} and preceding plots are fairly secure. However the absolute calibration of metallicities measured for extragalactic systems remains uncertain. Therefore the relationship to solar metallicity remains somewhat tenuous, as does the alignment of the three methods.

Another concern is that the ratios of CO, H$_2$, and SFR vary with time. On small scales this shows as offsets between H$\alpha$, a tracer of recent star formation, and CO, and induces large scatter in the respective ratios \citep[e.g.,][]{Schruba2010, Onodera2010} which can be linked to the evolution of star-forming regions \citep{Kawamura2009}. Numerical simulations even suggest that the ratios may never be constant inside a cloud because chemical equilibrium is not reached during most or all of a cloud's lifetime \citep{Glover2010, Glover2011, Shetty2011a, Shetty2011b, Feldmann2011a}. On galaxy scales however \citet{Pelupessy2009} and \citet{Papadopoulos2010} suggest that the ratios are roughly constant after dynamical equilibrium between ISM phases and stars is established ($t \sim 1$~Gyr). Only during (early) times of strong galaxy evolution, when the ISM phases and star formation are out of equilibrium, do larger deviations between CO, H$_2$, and SFR occur ($t \lesssim 0.2-0.3$ Gyr); gas-rich and/or low-metallicity galaxies can show strong periodic variations throughout their evolution. Such variations are strongest in the smallest dwarf galaxies ($M_{\rm B} > -15$) and are less common and weaker in (more) massive dwarfs and spirals \citep{Lee2009a}. Some galaxies of our literature sample experience a current starburst (e.g., NGC 2366, NCG 4449, and NGC 5253) or are in a post-starburst phase (e.g., NGC 1569). These bursts can last for a few 100 Myr \citep{McQuinn2010} and may show a strong time evolution in the brightness of their molecular gas and star formation tracers: starting with being bright in CO, followed by a phase being bright in CO and IR (a sign of embedded star formation), and finally being bright in IR and FUV \citep[sensitive to stellar populations of age $\lesssim 100$ Myr;][]{Salim2007}. Our literature samples, including SINGS, LVL, THINGS, and HERACLES have often an implicit or explicit bias to select IR-bright galaxies, and thus actively star-forming systems which means that we infer a high $\alpha_{\rm CO}$. Robust volume-limited surveys or otherwise unbiased samples are needed to remedy this.

\section{Summary}
\label{summary}

This paper presents sensitive maps of $^{12}\text{CO}~J=2-1$ emission for 16 nearby star-forming dwarf galaxies from the HERACLES survey \citep[for a first presentation of a subsample of our galaxies see][]{Leroy2009a}. Thanks to the large area covered ($\sim 2-5~R_{25}$) and high linear resolution of $\sim 250$~pc at the average target distance of $D=4$~Mpc, we can sample our targets by $10-1000$ resolution elements.

We apply the stacking techniques developed in \citet{Schruba2011} to perform the most sensitive search for CO emission in low-metallicity galaxies across the entire star-forming disk. We search for CO emission on three spatial scales: individual lines-of-sight, stacking over IR-bright regions indicating embedded star-formation and thus regions likely to contain molecular gas, and stacking over entire galaxies. Our point source sensitivity is $L_{\rm CO\,2-1} \sim 2 \times 10^4$ \Kkmperspc, sufficient to detect a CO-bright cloud with luminosity comparable to Orion~A or the brightest cloud in the SMC but at distance $D=4$~Mpc. When stacking over the entire galaxy our data have sufficient sensitivity to detect the LMC at $D=4$~Mpc; but not the SMC. We detect 5 galaxies in CO with total CO luminosities of $L_{\rm CO\,2-1} = 3 - 28 \times 10^6$ \Kkmperspc. The other $11$ galaxies remain undetected in CO even in the stacked images and have $L_{\rm CO\,2-1} \lesssim 0.4 - 8 \times 10^6$ \Kkmperspc.

We combine our dwarf galaxy sample with a large sample of spiral galaxies from the literature to study the relations between $L_{\rm CO}$, $M_{\rm B}$, and metallicity. We find that dwarf galaxies with metallicities $Z / Z_\odot \approx 1/2 - 1/10$ have $L_{\rm CO}$ of $2-4$ orders of magnitude smaller than massive spiral galaxies with $Z / Z_\odot \sim 1$ and that their $L_{\rm CO}$ per unit $L_{\rm B}$ is $1-2$ orders of magnitude smaller. Dwarf galaxies are thus significantly fainter in CO than a simple linear scaling with galaxy mass would suggest.

We also compare $L_{\rm CO}$ with tracers of recent star formation (FUV and 24\micron\ intensity) and find that $L_{\rm CO}$ per unit SFR is $1-2$ orders of magnitude smaller in dwarf galaxies as compared to massive spiral galaxies. The low $L_{\rm CO} / \text{SFR}$ ratios in dwarf galaxies may either indicate intrinsically small H$_2$ masses coupled with high star formation efficiencies or that CO emission becomes an increasingly poor tracer of H$_2$. The two are degenerate, however, following the arguments of recent observational studies of the dust-inferred gas content \citep{Leroy2011, Bolatto2011} and theoretical studies of the SFR-H$_2$ dependence \citep{Krumholz2011, Glover2012} we argue that the latter, i.e., significant changes in the CO-to-H$_2$ conversion factor, $\alpha_{\rm CO}$, at low metallicity are the dominant driver.

To estimate $\alpha_{\rm CO}$ and study its metallicity dependence we apply a method recently also used by \citet{Genzel2012} which assumes the conversion of H$_2$ to stars to be constant and infer H$_2$ masses and $\alpha_{\rm CO}$ values by scaling the observed total SFRs. We assume an H$_2$ depletion time of $\tau_{\rm dep} = M_{\rm H2} / \text{SFR} = 1.8$ Gyr, the average value found for massive spirals in the HERACLES sample \citep{Bigiel2008, Leroy2008, Schruba2011}. With this assumption we derive $\alpha_{\rm CO}$ values for dwarf galaxies with $Z / Z_\odot \approx 1/2 - 1/10$ more than one order of magnitude larger than those found in massive spiral galaxies with solar metallicity. This strong increase of $\alpha_{\rm CO}$ at low metallicity is consistent with previous studies, in particular those of Local Group dwarf galaxies which model dust emission to constrain H$_2$ masses \citep{Leroy2011, Bolatto2011}. Even though it is difficult to parametrize the dependence of $\alpha_{\rm CO}$ on metallicity given the currently available data the results suggest that CO is increasingly difficult to detect at lower metallicities. This has direct consequences for the detectability of star-forming galaxies at high redshift which presumably have on average sub-solar metallicity.

\acknowledgements

We thank the teams of SINGS, LVL, and {\em GALEX} NGS for making their outstanding data sets available. The work of A.S. was supported by the Deutsche Forschungsmeinschaft (DFG) Priority Program 1177. This research made use of the NASA/IPAC Extragalactic Database (NED), which is operated by the JPL/Caltech, under contract with NASA, NASAs Astrophysical Data System (ADS), and the HyperLeda catalog, located on the Worldwide Web at http://www.obs.univ-lyon1.fr/hypercat/intro.html.

\end{document}